# Methodology for Capacity Credit Evaluation of Physical and Virtual Energy Storage in Decarbonized Power System

Ning Qi[1*], Peng Li[2], Lin Cheng[1], Ziyi Zhang[2], Wenrui Huang[1], Weiwei Yang[2]

*1. State Key Laboratory of Control and Simulation of Power Systems and Generation Equipment, Tsinghua University, Beijing, 100084, China;*

*2. China Three Gorges Renewables (Group) Co., Ltd, Tongzhou District, Beijing, 101100, China*

*Abstract*—Energy storage (ES) and virtual energy storage (VES) are key components to realizing power system decarbonization. Although ES and VES have been proven to deliver various types of grid services, little work has so far provided a systematical framework for quantifying their adequacy contribution and credible capacity value while incorporating human and market behavior. Therefore, this manuscript proposed a novel evaluation framework to evaluate the capacity credit (CC) of ES and VES. To address the system capacity inadequacy and market behavior of storage, a two-stage coordinated dispatch is proposed to achieve the trade-off between day-ahead self-energy management of resources and efficient adjustment to real-time failures. And we further modeled the human behavior with storage operations and incorporate two types of decision-independent uncertainties (DIUs) (operate state and self-consumption) and one type of decision-dependent uncertainty (DDUs) (available capacity) into the proposed dispatch. Furthermore, novel reliability and CC indices (e.g., equivalent physical storage capacity (EPSC)) are introduced to evaluate the practical and theoretical adequacy contribution of ES and VES, as well as the ability to displace generation and physical storage while maintaining equivalent system adequacy. Exhaustive case studies based on the IEEE RTS-79 system and real-world data verify the significant consequence (10%-70% overestimated CC) of overlooking DIUs and DDUs in the previous works, while the proposed method outperforms other and can generate a credible and realistic result. Finally, we investigate key factors affecting the adequacy contribution of ES and VES, and reasonable suggestions are provided for better flexibility utilization of ES and VES in decarbonized power system.

## I. INTRODUCTION

### A. Motivation

Ensuring system adequacy is a primary function of power systems and is concerned with the existence of sufficient facilities within the system to satisfy demand [1]. System adequacy historically relied on fossil fuel power plants, but they are gradually replaced by the increasing penetration of renewable energy sources (RES) due to the energy crisis and environmental challenges [2]-[3]. RES can take responsibility for adequacy contribution but are highly subject to their inherent stochastic properties, thus stimulating the capacity market to target ES and demand response (DR) resources to provide the required flexibility and enhance adequacy performance [4]. The significant reliability improvement effects have been verified in previous works with the integration of these emerging resources. For instance, the reliability of a radial distribution system is enhanced by the optimal placement and sizing of energy storage systems [5]. Ref. [6] proposed a novel virtual energy storage (VES) model for DR, and the reliability of the distribution system is improved by optimal dispatch of VES considering endogenous uncertainties. However, it is still critical for power system operators and capacity market designers to qualify whether and to what extent we could rely on these resources, i.e., how to determine the credible capacity value of these resources.

### B. Literature review

The initial purpose of the capacity value/capacity credit evaluation was to provide a fair basis of comparison between conventional and renewable generation in terms of their contribution to system reliability [7]. However, the rapid development of the integration of ES and DR/VES resources has driven the extension and enrichment of CC evaluation. In addition to the original approach using the effective load carrying capability (ELCC) [8], various CC indices are available in the literature to qualify the CC of ES and VES, e.g., equivalent firm capability (EFC) [7], equivalent conventional capacity (ECC) [9], and equivalent generation substituted (EGCS) [10]. While different from the evaluation methodology of a generator, CC evaluation of ES and DR/VES normally involves the adequacy-oriented optimization problem to determine the scheduling of these resources, thus their CC values are particularly affected by the underlying dispatch methods and modeling of these resources, which are mainly addressed in this manuscript.

The adequacy-oriented optimization problem with ES and DR/VES is normally modeled in two distinct ways, i.e., fixed dispatch and 2) greedy management. For fixed dispatch, the scheduling of ES and DR is deterministic or even fixed. For ins-

*Corresponding author:
Ning Qi, Tsinghua University; e-mail address: qn18@mails.tsinghua.edu.cn

**NOMENCLATURE**

**Abbreviations**

| | |
|---|---|
| CC | Capacity credit |
| CG, RG | Conventional generation, renewable generation |
| DERs | Distributed energy resources |
| DIUs, DDUs | Decision-dependent uncertainties, decision-independent uncertainties |
| DR | Demand response |
| ES, VES, GES | Energy storage, virtual energy storage, generic energy storage |
| LR | Load recovery |
| SoC, SoH | State of charge, state of health |
| SMCS | Sequential Monte Carlo Sampling |

**Decision Variables**

| | |
|---|---|
| $P_{\mathrm{c/d},i,t}^{\mathrm{GES}}$ | Charge and discharge power of GES at bus $i$ and time $t$ |
| $SoC_{i,t}^{\mathrm{GES}}$ | SoC of GES at bus $i$ and time $t$ |
| $RD_{i,t}^{\mathrm{GES}}$ | Response discomfort of GES at bus $i$ and time $t$ |
| $P_i^{\mathrm{PK}}$ | Peak load of one day at bus $i$ |
| $P_{i,t}^{\mathrm{LC}}$ | Power of load curtailment at bus $i$ and time $t$ |
| $P_{i,t}^{\mathrm{CG}}$, $P_{i,t}^{\mathrm{RG}}$ | Power of conventional and renewable generation at bus $i$ and time $t$ |
| $\theta_{i,t}$, $P_{ij,t}$ | Voltage phase angle of bus $i$ and the power flow in the line $ij$ |

**Parameters**

| | |
|---|---|
| $P_{\mathrm{c/d},i,\max}^{\mathrm{GES}}$ | Upper charge and discharge power bounds of GES at bus $i$ |
| $SoC_{i,t,\max/\min}^{\mathrm{GES, DIU}}$ | Upper and lower SoC bounds of GES at bus $i$ and time $t$ under DIUs |
| $SoC_{i,t,\max/\min}^{\mathrm{GES, DDU}}$ | Upper and lower SoC bounds of GES at bus $i$ and time $t$ under DDUs |
| $SoC_t^{\mathrm{GES, BS}}$ | Baseline SoC of GES at time $t$ |
| $\eta_{\mathrm{c/d},i}^{\mathrm{GES}}$ | Charge and discharge efficiency of GES at bus $i$ |
| $\varepsilon_i^{\mathrm{GES}}$ | Self-discharge rate at bus $i$ |
| $S_i^{\mathrm{GES, rated}}$, $S_i^{\mathrm{GES, AV}}$ | Rated and available energy capacity of GES at bus $i$ |
| $\Delta t$, T | Time-step for dispatch |
| $P_{i,t}^{\mathrm{LD}}$ | Load power at bus $i$ and time $t$ |
| $R, C$ | Thermal resistance and capacity |
| $P_{\mathrm{c/d},t}^{\mathrm{GES, BS}}$ | Baseline power consumption of VES at time $t$ |
| $\omega_t^{\mathrm{CG/RG/ES/VES}}$ | On-off state or working/failure state of CG/RG/ES/VES at time $t$ |
| $P_t^{\mathrm{CG/RG, AV}}$, $C_t^{\mathrm{CG/RG, AV}}$ | Available power and capacity of CG/RG at time $t$ |
| $\lambda$, $\nu$ | The failure rate and repair rate of CG |
| $\alpha_t$ | Cumulative capacity degradation of the ES after the time $t$ |
| $L$ | The life circle of ES |
| $SoH_{\mathrm{initial}}$, $SoH_{\mathrm{end}}$ | Initial and end of SoH |
| $E_{\mathrm{c/d},t}^{\mathrm{ES}}$ | Charge and discharge energy of time $t$ |
| $a_g$, $b_h$ | Coefficients of incentive and discomfort DDUs impact level |
| $\gamma$ | Probability level of chance constraints |
| $\mu$, $\sigma$ | Mean and standard variance of the distribution |
| $\rho$ | Tradeoff weight over accumulated response power and SoC-based discomfort |
| $x_{ij}$, $P_{ij,\max}$ | Reactance and maximum power flow of line $ij$ |
| $C_{\mathrm{c/d},i}^{\mathrm{VES}}$, $\overline{C}^{\mathrm{VES}}$ | Charge/discharge capacity price, and maximum capacity price of VES at bus $i$ |
| $RC_t$ | Residual generation capacity at time $t$ |
| $\beta_i$ | Capacity share of each GES |
| N | Number of Monte Carlo scenarios |
| $c_t^{\mathrm{G}}$ | Day-ahead time-of-use (ToU) electricity price at time $t$ |

**Indices**

| | |
|---|---|
| $ERNS^{\mathrm{T}}$, $ERNS^{\mathrm{P}}$ | Theoretical and practical expected energy not served |
| EFC, ECC, EPSC | Equivalent firm, conventional generation, physical storage capacity |
| EGCS, ELCC | Equivalent generation capacity substituted, effective load carrying capability |

**Sets**

| | |
|---|---|
| $\Omega_{\mathrm{T}}$, $\Omega_{\mathrm{B}}$, $\Omega_{\mathrm{L}}$ | Sets of time series, buses and lines |
| $\boldsymbol{x}$, $\boldsymbol{y}$ | Sets of decision variables and stochastic parameters |

**Functions**

| | |
|---|---|
| $\boldsymbol{g}(\cdot)$, $\boldsymbol{h}(\cdot)$ | DDU effects functions associated with incentive and response discomfort |
| $\mathcal{N}(\cdot)$, $\mathcal{LN}(\cdot)$ | Normal and lognormal distribution |
| $\mathbb{P}(\cdot)$, $\boldsymbol{E}(\cdot)$, $R(\cdot)$ | Probability, expectation and reliability function |

-tance, the ES scheduling is optimized in [10] based on a daily peak shaving problem, which addresses the utilization of ES in reducing peak load and expected energy not served (EENS). The impact of different penetration of responsive load on the adequacy of the distribution system is evaluated in [11] by using the fixed responsive load profiles as the negative inputs of load. Besides, the load recovery (LR) characteristic is further considered in and added to responsive load profiles [12]. However, the operations under fixed dispatch [10]-[12] are determined before evaluation simulation so they are failed to account for the impact of unforeseen fault events in generation and network, which will result in the underestimated CC value. While different from the fixed dispatch, greedy management is proposed in [13] that storage facilities remain fully charged at normal system state to ensure the maximum contribution to system reliability through their immediate discharge as soon as a power shortfall occurs. And a more greedy strategy is generated based on the hypothesis of perfect foresight of forced outage of generation [14], resulting in an optimistic dispatch of ES resources and overvalued CC value. Therefore, it should be pointed out that the existing dispatch method is either over-simplified or unrealistic since they overlook the treatment of capacity inadequacy caused by real-time faults events and peak load events. And this problem is also reported by the European Network of Transmission System Operators (ENTSO-E) [15] that an inherent limitation of the applied methodology is the treatment of real-time system events.

Another problem open to discuss is the rational modeling of ES and DR/VES within the dispatch problem, especially focusing on the available power & energy capacity affected by human and market behaviors. Previous works [10][14] normally regard the rated capacity as the available capacity of storage used for adequacy-oriented optimization. Nevertheless, both ES and DR/VES resources are generally controlled by market players and can not guarantee 100% available capacity all the time. This is stated by the Electric Power Research Institute (EPRI) that the availability of storage capacity may be subject to limitations in real-time operation [16], such as forced outages and day-ahead market schedules. And it is also pointed out by California independent system operator (CAISO) [17], that a great quantity of DR customers refuse to fulfill their capacity contribution and the theoretical CC of DR appeared to be over-counted. Therefore, the available capacity of ES and DR/VES is considered to be stochastic and time-varying in this manuscript, which is due to 1) baseline consumption and 2) various types of DIUs and DDUs. On one hand, both ES and DR/VES resources normally operate under self-energy management to fulfill local system assignments. For instance, ES is dispatched to correct the day-ahead forecast errors of wind generation or improve the bidding profit of the wind farm [18]. The baseline consumption of VES generally accounts for a substantial energy input of the VES model [19]. Thus, the residual capacity for adequacy enhancement is much less than the rated capacity. Furthermore, DIUs and DDUs are inevitably involved in the operation of ES and DR/VES [6] [19]. Uncertainties in ES resources mainly stem from forced outage [20] and capacity degradation [21], which are widely investigated in previous works but are by far not included in the CC evaluation. While the uncertainties affected by incentive and discomfort are mostly considered for DR/VES operations. The responsivity of DR driven by price or incentive is generally stochastic and modeled as an exogenous probabilistic distribution in [22]-[23]. The response time availability of DR is constructed in [24] with the two-state Markov chain. It should be pointed out that the uncertainties in DR/VES are normally modeled as a fixed and exogenous probability distribution that can be fully determined before the CC evaluation. However, some other stochastic properties (e.g., temperature preference) may be affected by the control and price signals. These are generally overlooked in the work above [22]-[24], this type of stochasticity is called DDUs. Failing to incorporate such a dependent and dynamic nature of DR/VES is likely to cause a misestimation of its reliability benefits [25]. And ref. [6] verify the benefit of incorporating DDUs into the operations of VES, which will reduce the real-time unavailability of VES and hence improve practical performance. Thus, it is significant to incorporate DDUs within CC evaluation and reduce their bad consequences via an effective optimization approach.

### C. Contributions

To fill in the research gap in both modeling and evaluation methodologies, this manuscript addresses the capacity credit evaluation of ES and VES in the transmission power system with special consideration of real-time failures and available capacity of storage, thus providing a credible and realistic evaluation for their adequacy contribution. Specifically, the main contributions of this manuscript are threefold:

(i) Treatment to failures: The two-stage coordinated dispatch is proposed within CC evaluation to achieve the trade-off between self-energy management of ES and VES resources and efficient adjustment to real-time failures. The proposed method initially generates the day-ahead self-energy management strategy (e.g., energy arbitrage for ES, and self-consumption for VES) under the normal system state, while the real-time corrective dispatch of storage manages to minimize the loss of load under the emergency state. Compared with the existing fixed dispatch [10] and greedy management [13] methods, the proposed method guarantees a compromised and realistic value.

(ii) Treatment to the available capacity of storage: We incorporate various types of DIUs and DDUs within the coordinated dispatch of storage. Two types of DIUs (operate state and self-consumption) and one type of DDUs (available capacity) are detailed modeled, while the inherent difference exists in the uncertainties description between ES and VES. The case study simulates the dynamic and available capacity of storage with the proposed method and verifies the response unavailability (during DR) and load rebound (after DR) by the consequence (i.e., around 10%-70% overestimated CC value) of overlooking DIUs and DDUs in the previous works.

(iii) Novel indices and findings of CC evaluation: Novel reliability and CC indices are proposed to evaluate the practical adequacy performance and the ability of VES in replacement of ES. And multi-factors affecting CC evaluation of ES and VES are analyzed to improve the adequacy contribution, including dispatch method, uncertainties, rated power and energy capacity, penetration of RES, correlation factor, DDUs level, etc. The case study shows that the proposed method substantially outperforms previous ones in terms of practical adequacy and economic performance due to (1) incorporating DIUs and DDUs in the coordinated dispatch and (2) obtaining additional profits from self-energy management. Moreover, general suggestions and conclusions are provided for the capacity market and decarbonized power system.

### D. Paper organization

The remainder of the paper is organized as follows. The novel methodology of CC evaluation is presented in Section II. Numerical studies based on real-world data are proposed in Section III to illustrate the comparative performance. The application of the proposed method in the decarbonized power system is discussed in Section VI. The status quo and suggestion of the capacity market are proposed in Section II. Finally, conclusions are summarized in Section VII.

## II. NOVEL METHODOLOGY OF CAPACITY CREDIT EVALUATION OF PHYSICAL AND VIRTUAL ENERGY STORAGE

Power system adequacy is guaranteed by the power supply of various resources, which are manifested in four kinds: conventional generation (CG), renewable generation (RG), ES, and DR. While, some of the DR resources have the attributes and abilities of ES resources, hence motivating the term "virtual energy storage" (VES) [6]. We use VES instead of DR for modeling and description in the following. In this section, the essential parts of the evaluation methodology, i.e., evaluation framework, uncertainty description and modeling, two-stage coordinated dispatch method, and definition of capacity credit indices are detailed as follows.

### A. Evaluation Framework

The general procedures for CC evaluation of ES and VES are illustrated in Fig. 1, and the detailed explanation for each step in Fig. 1 is presented below.

**Step 1:** Data inputs involve 1) topology and electrical parameters of the network; 2) capacity of generation and historical profile of RG; 3) load profile from historical data or used in reliability test system; 4) reliability parameters of adequacy resources; 5) model parameters of ES and VES.

**Step 2:** Sequential Monte Carlo simulation (SMCS) is used to randomly generate state-duration time series of adequacy resources. Herein, we choose SMCS as a reliability analysis tool because unlike other techniques (such as the analytical method [26] and normal Monte Carlo simulation [27]), SMCS examines the reliability indices by generating an artificial history of failure and repair events in a chronological manner. Thus, it can maintain inter-temporal information about system behaviors. Within SMCS, the forced outage rate (FOR) is used to determine the binary state of lines of the network and generations. The stochastic characteristics of RG and load can be obtained via data-driven methods. The description and simulation of the stochasticity of the resources will be detailed in Section II. B.

**Step 3:** Two-stage coordinated dispatch of ES and VES is generated by Algorithm A with special consideration of the real-time failures and available capacity of the resources. The coordinated dispatch involves day-ahead self-energy management (normal state) and real-time corrective dispatch (emergency state). For day-ahead operations, ES and VES will follow the self-energy management strategy. While, if the emergency state emerges in real-time operation, ES and VES will adjust the day-ahead strategy and discharge more to minimize the loss of load. The dispatch method will be detailed in Section II. C.

**Step 4:** After processing **Step 1-3**, the state of the network and the available power output of adequacy resources are determined and ready for load curtailment optimization. By using Algorithm $B_1$ and $B_2$, load curtailment is checked repeatedly over each time period $t$. The frequency and quantity of the loss of load event are counted until the convergence of the algorithm, and reliability indices are finally counted. Since the existing reliability indices are massive, the expected energy not served (EENS) and loss of load probability (LOLP) are normally chosen as the reference reliability indices.

**Step 5:** After calculating the reliability of the reference and test system, we can calculate the capacity credit of ES or VES by the replacement of generation, load, or storage in the reference system, The definition of CC metrics will be detailed in Section II. D.

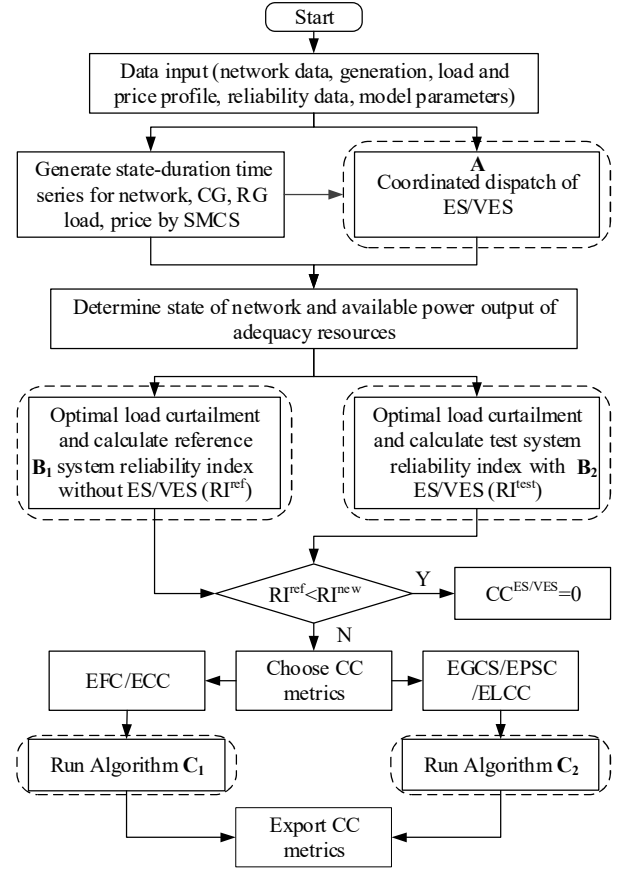

Fig. 1. Flowchart of capacity credit evaluation

### B. Uncertainty description and modeling

In this subsection, the uncertainty description and modeling of different resources considered within CC evaluation are modeled as follows.

**(a) Conventional Generation**

CG is generally controlled by the system operator as the main adequacy supply of the system. And it can adjust the power supply by the system dispatch commands, but its real-time operation is affected by its forced outage. And the uncertain state (i.e., normal state and fault state), can be modeled as a two-state Markov Chain:

$$f(\omega_t) = \begin{cases} \dfrac{v}{\lambda + v} = \dfrac{\text{MTTR}}{\text{MTTR} + \text{MTTF}}, & \omega_t = 1 \\ \dfrac{\lambda}{\lambda + v} = \dfrac{\text{MTTF}}{\text{MTTR} + \text{MTTF}}, & \omega_t = 0 \end{cases} \quad (1)$$

Where the state of generation is defined as $\omega_t$. Subscript $t$ defines the time interval. The failure rate and repair rate are defined as $\lambda$ and $v$. These two parameters are normally replaced with the mean time to failure (MTTF) and mean time to repair (MTTR) in practice and the two definitions are the same in essence. This uncertainty is a DIU and can be simulated within SMCS.

**(b) Renewable Generation**

Renewable generation (RG) (e.g., wind and solar generation) is more uncontrollable and stochastic than CG, and the uncertainties are DIUs and manifested in both operation state and available capacity. The modeling of the operation state of RG is consistent with CG using a two-state Markov Chain as Eq. (1). While the available capacity of RG is driven by environmental factors (e.g., wind speed, solar radiation, etc.). And its stochastic power output is commonly simulated using mean capacity factor from historical data [28], autoregressive moving average (ARMA) [29], or generative adversarial networks [30]. Based on the above models, the available power output of CG or RG unit can be expressed as follows:

$$P_t^{CG/RG,AV} = \omega_t^{CG/RG} C_t^{CG/RG,AV} \quad (2)$$

**(c) Energy Storage**

Energy storage (ES) is normally controllable, but its real-time operation will be affected by the forced outage and capacity degradation. While various kinds of energy storage exist (e.g., battery, pumped hydro storage, etc.), only battery energy storage is specially discussed in this study for illustration. The DIUs of ES are concluded and listed as follows:

(i) DIU: operation state. To be simplified, the operation state of ES is modeled as a DIU and the on-off state $\omega_t^{ES}$ can be modeled as a Bernoulli distribution:

$$f(\omega_t^{ES}) = \begin{cases} 1 - \text{FOR}^{ES} & \omega_t^{ES} = 1 \\ \text{FOR}^{ES} & \omega_t^{ES} = 0 \end{cases} \quad (3)$$

Where the forced outage rate of ES is defined as $\text{FOR}^{ES}$ and can be obtained from industrial standards or historical data.

(ii) DIU: baseline consumption. The baseline consumption is generated from the self-energy management strategy of ES, while it differs across different sectors. For ES bundled with

RG, the baseline strategy is to minimize the day-ahead forecast error and improve the bidding profit of RG. While for ES distributedly located in the demand sector, the baseline strategy is to maximize the profit via energy arbitrage. The baseline consumption can be simulated with the operation profile of RG and price, and it will be detailed in Section II. C.

(iii) DDU: available capacity affected by capacity degradation. The capacity degradation includes the cycle and calendar degradation, and the cycle degradation is solely considered in this work since the cycle degradation accounts for a substantial part of capacity degradation compared with the calendar degradation. The capacity degradation of different types of batteries depends on the rated energy capacity, the equivalent cycle number [31], and charge/discharge actions, thus the capacity degradation is complex and decision-dependent, and the available capacity can be calculated by:

$$\alpha_t = \alpha_{t-1} + \kappa \times \frac{E_{c,t}^{ES} + E_{d,t}^{ES}}{S^{ES,rated}} \times \frac{1}{L^{SoH_{inital} - SoH_{end}}} \quad (4)$$

$$S_t^{ES,AV} = S^{ES,rated} \times [SoH_{initial} - (SoH_{initial} - SoH_{end})\alpha_t] \quad (5)$$

Where $\alpha_t$ defines the cumulative capacity degradation of the battery after the time interval $t$, expressed as a percentage. $L$ and $L^{SoH_{inital} - SoH_{end}}$ represent the life circle and the equivalent cycle number of battery that capacity reduces from the beginning of use (expressed as $SoH_{inital}$) to the end of its life (expressed as $SoH_{end}$) due to the charge actions $E_{c,t}^{ES}$ and discharge actions $E_{d,t}^{ES}$. $SoH$ defines the state of health. $S_t^{ES,av}$ is the available capacity that is reduced from the rated capacity $S^{ES,rated}$. $\kappa$ is a probabilistic parameter ($0 < \kappa < 1$).

**(d) Virtual Energy Storage**

Virtual energy storage (VES) shares the similarity with ES and can be modeled in the same form as the ES model. However, VES is more stochastic and time-varying compared with ES, and we present the uncertainty description and modeling of VES in our recent work [6]. The DIUs and DDUs of VES are concluded and listed as follows:

(i) DIU: operation state. VES resources can only respond to DR commands when they operate in an on-state or normal state. Thus, they can not maintain 100% reliable response all the time which is due to customers' consumption behavior. This stochasticity is a DIUs and the on-off state $\omega_t^{VES}$ can be modeled as a Bernoulli distribution:

$$f(\omega_t^{VES}) = \begin{cases} p_t^{VES} & \omega_t = 1 \\ 1 - p_t^{VES} & \omega_t = 0 \end{cases} \quad (6)$$

Where on-state probability $p_t^{VES}$ can be obtained from historical data via a data-driven approach.

(ii) DIU: self-consumption. Since VES resources are not fully controlled by the system operator, day-ahead self-consumption accounts for a substantial part of overall energy actions besides the demand response to the grid. The uncertain baseline consumption can be determined by historical data. Without loss of generality, we use the data-driven results in [32] to model these DIUs, where baseline power consumption obeys

Lognormal distribution and SoC bounds obey Normal distribution:

$$P_{c/d,t}^{VES,DA} \sim \mathcal{LN}(\mu_{p_{c/d,t}^{VES,DA}}, \sigma_{p_{c/d,t}^{VES,DA}}) \quad (7)$$

$$SoC_{t,min/max}^{VES} \sim \mathcal{N}(\mu_{SoC_{t,min/max}^{VES}}, \sigma_{SoC_{t,min/max}^{VES}}) \quad (8)$$

Where $P_{c/d,t}^{VES,BS}$ defines the day-ahead baseline power consumption, and its mean and standard deviation are denoted as $\mu_{p_{c/d,t}^{VES,BS}}$ and $\sigma_{p_{c/d,t}^{VES,BS}}$. The upper and lower bounds of SoC are defined as $SoC_{t,min/max}^{VES}$.

(3) DDU: available capacity affected by incentive and discomfort effect. The capacity of VES to respond is an important decision-dependent uncertainty to consider. We introduce the available SoC bounds for description. Normally, the SoC bounds are fixed to be "0" and "1" for physical energy storage as illustrated as blue lines in Fig. 2, while the SoC bounds of VES are associated with DIUs and time-varying properties as described in DIUs (ii), which are marked with green rainbow lines. Further incorporating DDU, the decisions and price signals will affect the SoC bounds, which come as a trade-off between inconvenience costs (i.e., the discomfort they sustained during DR events) and the expected earnings (i.e., incentive payment for DR). Thus, SOC bounds of VES will be expanded or contracted based on incentive or discomfort, which are marked with red rainbow lines. The specific structure of DDUs is given in Eq. (9-13), but it can be a general form which is proved in our previous work [19].

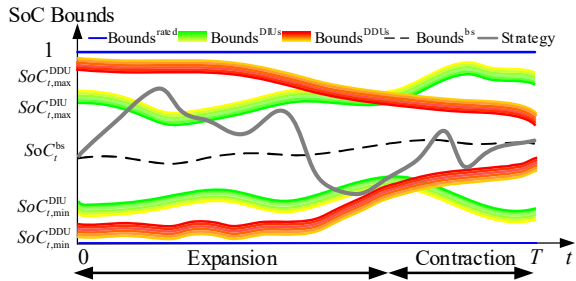

Fig. 2. Visualization of DIUs and DDUs in SoC bounds of VES

$$SoC_{t,max}^{VES,DDU} = SoC_{t,max}^{VES,DIU}(1 + \boldsymbol{g}(\mu_{\underline{g}}, \sigma_{\underline{g}}))(1 - \boldsymbol{h}(\mu_{\bar{h}}, \sigma_{\bar{h}})) \quad (9)$$

$$SoC_{t,min}^{VES,DDU} = SoC_{t,min}^{VES,DIU}(1 - \boldsymbol{g}(\mu_{\underline{g}}, \sigma_{\underline{g}}))(1 + \boldsymbol{h}(\mu_{\underline{h}}, \sigma_{\underline{h}})) \quad (10)$$

$$\mu_{\bar{g}} = a_g C_{c,i}^{VES} / \overline{C}^{VES}, \ \mu_{\underline{g}} = a_g C_{d,i}^{VES} / \overline{C}^{VES} \quad (11)$$

$$\mu_{\bar{h}} = b_h RD_t, \ \mu_{\underline{h}} = b_h RD_t \quad (12)$$

$$RD_t = \rho \sum_{\tau=1}^{t} (P_{c,\tau}^{VES,RT} / P_{c,max}^{VES} + P_{d,\tau}^{VES,RT} / P_{d,max}^{VES}) / T \\ + (1-\rho) \left| SoC_t^{VES,RT} - SoC_t^{VES,DA} \right| \quad (13)$$

Where the SoC bounds are denoted with DDUs and DIUs superscript. $\boldsymbol{g}$ and $\boldsymbol{h}$ are the distribution related to incentive and discomfort effects, which will increase and decrease the capacity flexibility of VES. And it can be broad among Lognormal distribution, Beta distribution, etc. μ and σ are the mean and standard variance of the distribution. $a_g$ and $b_h$ are coefficients of DDUs that reveal the impact level of different VES. $C_{c/d,t}^{VES}$ and $\overline{C}^{VES}$ are the charge/discharge capacity price, and maximum capacity price. $RD_t$ is the response discomfort composed of disutility and SoC deviation to the baseline value $SoC_t^{VES,DA}$, and $\rho$ ($0 \le \rho \le 1$) is the weight over these two items. Different types of VES bear different values of $\rho$. $SoC_t^{VES,RT}$ and $P_{c/d,t}^{VES,RT}$ are the real-time SoC, charge/discharge power strategy of VES. $P_{c,max}^{VES}$ and $P_{d,max}^{VES}$ are the power bounds of VES. $T$ defines the whole dispatch time period.

**(e) Load**

Load is generally uncontrollable and stochastic, and the certain power $P_t^{LD}$ is a DIUs and can be simulated via SMCS from historical data or standard case data.

The uncertainty description of five different resources is summarized in Table I. It should be mentioned that the uncertainties of CG, RG, and load are all DIUs and simulated with SMCS. And the uncertainty description of ES and VES are homologous, which motivates the new item "generic energy storage" (GES). We will use GES for unified modeling in the dispatch method. To be tractable, we newly introduce the chance-constrained optimization method to deal with uncertainties (3) of GES in this paper.

TABLE I Uncertainty Description of Different Resources

| Resources | Uncertainty Type | Uncertainty Manifestation | Simulation Method |
|---|---|---|---|
| CG | DIUs | Operate State | SMCS |
| RG | DIUs | Operation State & Power | SMCS |
| ES/VES | DIUs | Operation State Self Consumption | SMCS |
| | DDUs | Available Capacity | CCO |
| Load | DIUs | Power | SMCS |

**C. Two-stage coordinated dispatch method**

In this subsection, the modeling and scheduling principle of the proposed coordinated dispatch method is provided, while compared with (i) fixed dispatch and (ii) greedy management.

**(i) Fixed Dispatch**

The fixed dispatch of GES is typically conducted during summer or winter peak load days. And the dispatch strategy is determined exogenously to the SMCS, without considering the adjustment to real-time failures and the available capacity of GES. The formulation of fixed dispatch is described in Eq (14-20). The objective is to minimize the peak load at each bus $P_i^{PK}$. The relationship between net load and peak load power is given as the constraint (15). Constraints (16-17) limit the charge/discharge power of GES. Constraint (18) limits the state of charge (SoC) of GES, and constraint (19) ensures a sustainable energy state for the GES over time. Constraint (20) defines the relationship among charge power, discharge power, and SoC.

**Objective Function:**

$$\min P_i^{\text{PK}} \tag{14}$$

**Constraints:** $\forall t \in \Omega_{\text{T}}$ , $\forall i \in \Omega_{\text{B}}$

$$P_{i,t}^{\text{LD}} + P_{\text{c},i,t}^{\text{GES}} - P_{\text{d},i,t}^{\text{GES}} - P_{i,t}^{\text{CG/RG,AV}} \leq P_i^{\text{PK}} \tag{15}$$

$$0 \leq P_{\text{c},i,t}^{\text{GES}} \leq P_{\text{c},i,\max}^{\text{GES}} \tag{16}$$

$$0 \leq P_{\text{d},i,t}^{\text{GES}} \leq P_{\text{d},i,\max}^{\text{GES}} \tag{17}$$

$$SoC_{i,\min}^{\text{GES}} \leq SoC_{i,t}^{\text{GES}} \leq SoC_{i,\max}^{\text{GES}} \tag{18}$$

$$SoC_{i,T}^{\text{GES}} = SoC_{i,0}^{\text{GES}} \tag{19}$$

$$SoC_{i,t+1}^{\text{GES}} = (1 - \varepsilon_i^{\text{GES}}) SoC_{i,t}^{\text{GES}} + \frac{(\eta_{\text{c},i}^{\text{GES}} P_{\text{c},i,t}^{\text{GES}} - P_{\text{d},i,t}^{\text{GES}} / \eta_{\text{d},i}^{\text{GES}}) \Delta t}{S_i^{\text{GES,rated}}} \tag{20}$$

Where the sets of time periods and buses are defined as $\Omega_{\text{T}}$ and $\Omega_{\text{B}}$. The maximum charge and discharge power of ES/VES are given by $P_{\text{c},i,\max}^{\text{GES}}$ and $P_{\text{d},i,\max}^{\text{GES}}$ . $\varepsilon_i^{\text{GES}}$ defines the self-discharge rate of ES/VES. Charge and discharge efficiency are defined as $\eta_{\text{c},i}^{\text{GES}}$ and $\eta_{\text{d},i}^{\text{GES}}$ . $\Delta t$ is the time interval.

**(ii) Greedy Management**

The greedy management is to address the loss of load risks, and strategy is determined within the SMCS process, in response to real-time failures in the generations. The operation state of GES is changeable due to the changes in the operation state of the system.

(1) Normal state: GES will remain fully charged by the residual generation capacity $RC_t$ under the normal state (without loss of load). The residual generation capacity is calculated over the whole system and given by Eq. (21). The real-time charge power is distributed by the capacity of each GES and is subject to the inherent rated power and capacity limit. And it is calculated by Eq. (22). $\varphi_i$ defines the capacity share of each GES.

$$RC_t = \sum_i (P_{i,t}^{\text{CG/RG,AV}} - P_{i,t}^{\text{LD}}) \tag{21}$$

$$P_{\text{c},i,t}^{\text{GES}} = \min\{\varphi_i RC_t, P_{\text{c},i,\max}^{\text{GES}}, [SoC_{i,\max}^{\text{GES}} - (1 - \varepsilon_i^{\text{GES}}) SoC_{i,t-1}^{\text{GES}}] \\ S_i^{\text{GES,rated}} / \eta_{\text{c},i}^{\text{GES}} \Delta t\} \tag{22}$$

(2) Emergency state: GES will immediately discharge under the emergency state (capacity deficits and loss of load appear). For simplification, the DC-OPF is employed and the problem is formulated as (23-31). The objective function is to minimize the total load loss of load and expressed as the sum of load curtailment. The constraints (24-25) are limitations for the DC-OPF. Power balance is given as the constraint (26). The constraints (27-28) limit the power of CG and RG. The power of load curtailment is limited as the constraint (29). Constraints (30-31) limit the SoC and discharge power of GES.

**Objective Function:**

$$\min \sum_i P_{i,t}^{\text{LC}} \tag{23}$$

**Constraints:** $\forall t \in \Omega_{\text{T}}$ , $\forall i,j \in \Omega_{\text{B}}$ , $\forall ij \in \Omega_{\text{L}}$

$$P_{ij,t} = (\theta_{i,t} - \theta_{j,t}) / X_{ij} \tag{24}$$

$$-P_{ij,\max} \leq P_{ij,t} \leq P_{ij,\max} \tag{25}$$

$$P_{i,t}^{\text{CG/RG}} + P_{i,t}^{\text{LC}} + P_{\text{d},i,t}^{\text{GES}} = \sum_{ij \in \Omega_{\text{L}}} P_{ij,t} + P_{i,t}^{\text{LD}} \tag{26}$$

$$0 \leq P_{i,t}^{\text{CG}} \leq P_{i,t}^{\text{CG,AV}} \tag{27}$$

$$(1 - r_i) P_{i,t}^{\text{RG,AV}} \leq P_{i,t}^{\text{RG}} \leq P_{i,t}^{\text{RG,AV}} \tag{28}$$

$$0 \leq P_{i,t}^{\text{LC}} \leq P_{i,t}^{\text{LD}} \tag{29}$$

$$SoC_{i,\min}^{\text{GES}} \leq (1 - \varepsilon_i^{\text{GES}}) SoC_{i,t}^{\text{GES}} - \frac{P_{\text{d},i,t}^{\text{GES}} \Delta t}{\eta_{\text{d},i}^{\text{GES}} S_i^{\text{GES,rated}}} \tag{30}$$

$$0 \leq P_{\text{d},i,t}^{\text{GES}} \leq P_{\text{d},i,\max}^{\text{GES}} \tag{31}$$

Where $\Omega_{\text{L}}$ defines the set of lines. $\theta_{i,t}$ and $X_{ij}$ define the voltage phase angle and the reactance of the line. The power flow and its maximum value are denoted as $P_{ij,t}$ and $P_{ij,\max}$. The load curtailment power is denoted as $P_{i,t}^{\text{LC}}$. $r_i$ is the maximum curtailed rate of RG.

**(iii) Coordinated Dispatch**

The fixed dispatch and greedy management are oversimplified and unrealistic, which will be invalid in real-time operation. Instead, we proposed a coordinated dispatch framework of GES, which involves three operation states and transition of GES:

1) Normal state: GES will adhere to their day-ahead self-consumption under the normal state (without loss of load). While the baseline consumption will be different for GES of different sectors. Herein, we consider GES from four sections, i.e., ES bundled with RG (ES$^{\text{R}}$), household or distributed ES (ES$^{\text{D}}$), VES from thermostatically controlled load (VES$^{\text{T}}$), and VES from electric vehicle (VES$^{\text{E}}$). For ES$^{\text{R}}$ and ES$^{\text{D}}$, ES follows the energy arbitrage scheduling to maximum the profit in the market operation, and the objective function is formulated as (32). The constraints of ES are in line with constraints (16-20), while the constraints (4-5) are added for available capacity affected by capacity degradation. And the historical consumption data is employed to generate the baseline consumption profile of VES$^{\text{T}}$ and VES$^{\text{E}}$.

**Objective Function:**

$$\max \sum_t c_t^{\text{G}} (P_{\text{d},i,t}^{\text{ES,DA}} - P_{\text{c},i,t}^{\text{ES,DA}}) \tag{32}$$

Where $c_t^{\text{G}}$ is the day-ahead time-of-use electricity price.

(2) Emergency state: GES will immediately discharge under the emergency state (capacity inadequacy and loss of load appear). The operation under the emergency state is similar to the greedy management, however, we incorporate the baseline consumption and DDUs into dispatch, which will reflect the real-time available capacity of GES. The whole optimization is formulated as (36). The operational constraints involve the DDUs effect denoted as Eq. (10-13), the DC-OPF constraints denoted as Eq. (24-29), the power limits of GES denoted as Eq. (33) and modified SoC constraints (34-35). And the solution methodology of chance-constrained optimization under DDUs is provided in [19].

$$0 \leq P_{\text{d},i,t}^{\text{GES,RT}} \leq P_{\text{d},i,\max}^{\text{GES}} \tag{33}$$

$$\mathbb{P}(SoC_{i,\min}^{\text{GES,DDU}} \leq SoC_{i,t}^{\text{GES,RT}}) \geq 1 - \gamma \tag{34}$$

$$SoC_{i,t+1}^{\text{GES,RT}} = (1-\varepsilon_i^{\text{GES}})SoC_{i,t}^{\text{GES,RT}} - \frac{P_{d,i,t}^{\text{GES,RT}}\Delta t}{\eta_{d,i}^{\text{GES}}S_i^{\text{GES,AV}}} \qquad (35)$$

$$\min \sum_i P_{i,t}^{\text{LC}}$$

$$\text{s.t. Constraints (10-13, 24-29)} \qquad (36)$$

$$\text{Constraints (33-35)}$$

Where $1-\gamma$ is the confidence level of the chance constraints.

(3) Recovery state: GES will recover back to its baseline operation profile after the emergency state is over (without loss of load). This part of coordinated dispatch is rational since the LR effect is widely observed in the DR application [33] and the customers prefer to remain at the SoC level as usual after the DR terminates. If the real-time SoC is lower than the baseline one, the additional charge power of GES is calculated as:

$$P_{c,i,t}^{\text{GES,RT}} = \min\{P_{c,i,\max}^{\text{GES}}, [SoC_{i,t}^{\text{GES,BS}} - (1-\varepsilon_i^{\text{GES}})SoC_{i,t-1}^{\text{GES,RT}}]$$
$$S_i^{\text{GES,AV}}/\eta_{c,i}^{\text{GES}}\Delta t,\ \varphi_i RC_t\} \qquad (37)$$

While if the real-time SoC is higher than the baseline one, the discharge power of GES is calculated as:

$$P_{d,i,t}^{\text{GES,RT}} = \min\{P_{d,i,\max}^{\text{GES}}, [(1-\varepsilon_i^{\text{GES}})SoC_{i,t-1}^{\text{GES,RT}} - SoC_{i,t}^{\text{GES,BS}}]$$
$$S_i^{\text{GES,AV}}\eta_{d,i}^{\text{GES}}/\Delta t\} \qquad (38)$$

To further elaborate the processing and implementation of the proposed coordinated dispatch, key steps of the algorithm are summarized in **Algorithm A,** which involves both the day-

---

**Algorithm A:** Coordinated dispatch of GES

**Sep 1  Day-ahead dispatch under normal state.**

    **if** GES= $\text{ES}^{\text{R}}$ or $\text{ES}^{\text{D}}$
      **do** optimization with objective (32), constraints (16-20, 4-5)
    **else if** GES= $\text{VES}^{\text{T}}$ or $\text{VES}^{\text{E}}$
      **do** data-driven approach to obtain baseline consumption
    **end**

**Sep 2  Real-time dispatch under emergency and recovery state**

    **for** $t$=1:$T$
      **if** $RC_t < 0$  **(emergency state)**
        **do** chance-constrained optimization under DDUs (36)
      **else if** $SoC_{i,t}^{\text{GES,DA}} < SoC_{i,t}^{\text{GES,RT}}$ **(recovery state)**
        $P_{d,i,t}^{\text{GES,RT}} \leftarrow \min\{P_{d,i,\max}^{\text{GES}}, [(1-\varepsilon_i^{\text{GES}})SoC_{i,t-1}^{\text{GES,RT}} - SoC_{i,t}^{\text{GES,DA}}]$
        $S_i^{\text{GES,AV}}\eta_{d,i}^{\text{GES}}/\Delta t\}$
        $P_{c,i,t}^{\text{GES,RT}} \leftarrow 0$
      **else if** $SoC_{i,t}^{\text{GES,DA}} > SoC_{i,t}^{\text{GES,RT}}$ **(recovery state)**
        $P_{c,i,t}^{\text{GES,RT}} \leftarrow \min\{P_{c,i,\max}^{\text{GES}}, [SoC_{i,t}^{\text{GES,DA}} - (1-\varepsilon_i^{\text{GES}})SoC_{i,t-1}^{\text{GES,RT}}]$
        $S_i^{\text{GES,AV}}/\eta_{c,i}^{\text{GES}}\Delta t, \varphi_i RC_t\}$
        $P_{d,i,t}^{\text{GES,RT}} \leftarrow 0$
      **else if** $SoC_{i,t}^{\text{GES,DA}} = SoC_{i,t}^{\text{GES,RT}}$ **(normal state)**
        $P_{c,i,t}^{\text{GES,RT}} \leftarrow 0$ , $P_{d,i,t}^{\text{GES,RT}} \leftarrow 0$
      **end**
    **end**

**Sep 3  Return** day-ahead and real-time strategy $P_{c,d,i,t}^{\text{GES,BS}}$ and $P_{c,d,i,t}^{\text{GES,RT}}$

---

ahead dispatch and the coordinated real-time dispatch. And Fig. 3 compares the system and GES operation on the same typical day, applying the proposed method and the above two alternatives. The purple and grey areas represent two "loss of load and capacity inadequacy" events due to 1) the failure in CG and 2) peak load. It is observed that GES operations are inherently different under three dispatch methods. Although the loss of load during peak load time is all addressed in the above methods, the fixed dispatch overlooked the solutions to the generation failures, which fails to capitalize the flexibility of GES. Moreover, both fixed dispatch and greedy management retain the capacity of GES at a relatively high level (50%-100%), which overlooks the baseline consumption and the reduced available capacity from DIUs & DDUs. By contrast, the coordinated dispatch achieves the trade-off utilization of GES capacity between the day-ahead baseline consumption and real-time adequacy support, while maintaining the SoC of GES at a feasible and realistic level. However, the difference between the day-ahead and real-time operations of GES is not considered in the existing two dispatch methods.

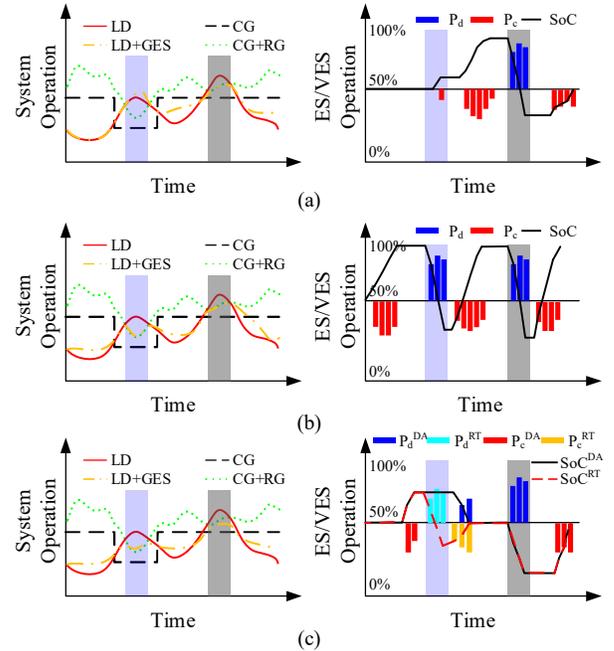

Fig. 3.  Visualization of system and GES operations under different dispatch methods: (a) fixed dispatch, (b) greedy management, (c) coordinated dispatch

### D. Reliability assessment with GES under DDUs

Reliability assessment is to evaluate the probability that a power system will perform its functions adequately without any failure within a stipulated period when subjected to normal operating conditions [1]. Previous reliability assessment aims to assess the reliability level of the system with exogenously stochastic factors from failures in the grid, the stochastic output of load, and RES, which is noted as reliability assessment under DIUs. And traditional reliability indices, e.g., loss of load probability (LOLP), power of load curtailment (PLC), and expected energy not served (EENS), etc., are commonly used to account for the ex-ante risk and overlook the ex-post risk of using these GES resources. As mentioned in our recent work [6], the practical operation of GES is inconsistent with the theoretical one, and overlooking the DDUs of GES will result in over-estimated reliability results or even worsen the system security level. Thus, without loss of generality, we focus on EENS and introduce the theoretical and practical reliability indices in Eq. (40). $X_{i,j} \mid \boldsymbol{y}$ represents the loss of load events without the unavailable response of adequacy resources under uncertainty $\boldsymbol{y}$. $Y_{i,j} \mid \boldsymbol{x}, \boldsymbol{y}$ represents the loss of load event due to the unavailable response of adequacy resources under decisions $\boldsymbol{x}$ and uncertainty $\boldsymbol{y}$. $\boldsymbol{E}$ is the expectation function. $N$ is the scenario number of SMCS. Moreover, the modified algorithm of reliability assessment under DDUs (B1 & B2) is actively demonstrated in the flowchart, i.e., Fig. 4. The optimal load curtailment for each sampled scenario is the essential part of reliability assessment, and the problem formulation and corresponding solution methodology are widely proposed in the previous works [6], [34], which will not be detailed in this paper.

$$
\begin{cases}
ERNS^{\mathrm{T}} = \sum_{i=1}^{N} \sum_{j=1}^{T} \boldsymbol{E}(X_{i,j} \mid \boldsymbol{y}) / NT \\
ERNS^{\mathrm{P}} = \sum_{i=1}^{N} \sum_{j=1}^{T} [\boldsymbol{E}(X_{i,j} \mid \boldsymbol{y}) + \boldsymbol{E}(Y_{i,j} \mid \boldsymbol{x}, \boldsymbol{y})] / NT
\end{cases}
\tag{39}
$$

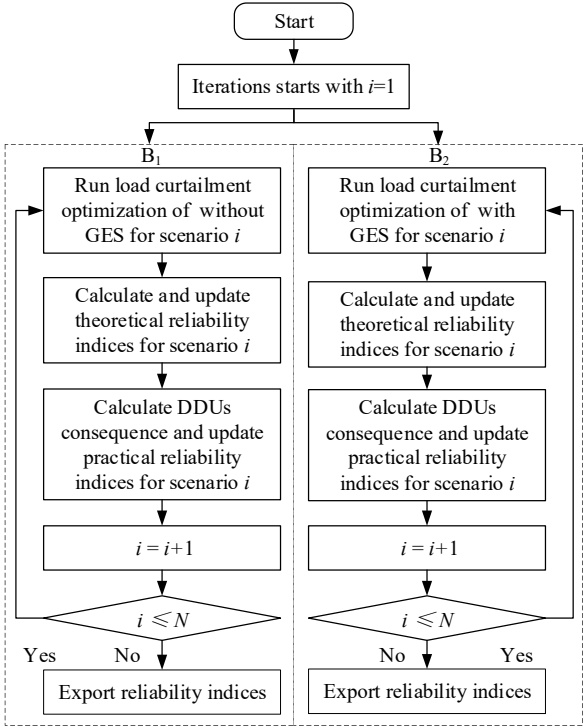

Fig. 4. Flowchart of Algorithm B1 & B2 for reliability assessment with GES

### E. Capacity credit indices

The capacity credit (CC) is to qualify the capacity of resources in terms of their contribution to system adequacy. In this subsection, we provide the classification of the CC indices and propose a novel CC index of VES for its replacement of ES. The CC indices can be clustered into two types:

i) Evaluation with adding capacity, such as EFC and ECC. And CC is calculated by increasing the capacity of generation of the original system to achieve reliability equivalence with the system supported by GES, which is expressed by Eq. (41). The reliability of reference and test system are denoted as $\mathrm{RI}^{\mathrm{ref}}$ and $\mathrm{RI}^{\mathrm{test}}$. $\mathrm{C}^{\mathrm{GES}}$ is the capacity of GES. $\mathrm{C}^{\mathrm{CG/RG,ADD}}$ is the additional capacity of CG/RG, representing the CC value.

$$
\begin{cases}
\mathrm{RI}^{\mathrm{ref}} = \boldsymbol{R}[\mathrm{C}^{\mathrm{CG/RG,AV}} + \mathrm{C}^{\mathrm{CG/RG,ADD}}] \\
\mathrm{RI}^{\mathrm{test}} = \boldsymbol{R}[\mathrm{C}^{\mathrm{CG/RG,AV}} + \mathrm{C}^{\mathrm{GES}}] \\
\mathrm{RI}^{\mathrm{ref}} = \mathrm{RI}^{\mathrm{test}}
\end{cases}
\tag{40}
$$

Herein, the equivalent physical storage capacity (EPSC) is newly proposed to qualify the capacity of VES in equivalence with physical energy storage. The definition is given by Eq. (41). And $\mathrm{C}^{\mathrm{RP}}$ is the replacing capacity of ES.

$$
\begin{cases}
\mathrm{RI}^{\mathrm{ref}} = \boldsymbol{R}[\mathrm{C}^{\mathrm{CG/RG,AV}} + \mathrm{C}^{\mathrm{RP}}] \\
\mathrm{RI}^{\mathrm{test}} = \boldsymbol{R}[\mathrm{C}^{\mathrm{CG/RG,AV}} + \mathrm{C}^{\mathrm{VES}}] \\
\mathrm{RI}^{\mathrm{ref}} = \mathrm{RI}^{\mathrm{test}}
\end{cases}
\tag{41}
$$

ii) Evaluation with replacing capacity, such as ELCC and EGCS. And CC is calculated by decreasing or replacing the existing load or generation in the system, which is expressed as Eq. (42). $\mathrm{LD}^{\mathrm{RP}}$ and $\mathrm{C}^{\mathrm{RP}}$ are the replacing capacity of load and generation, representing the CC value.

$$
\begin{cases}
\mathrm{RI}^{\mathrm{ref}} = \boldsymbol{R}[\mathrm{C}^{\mathrm{CG/RG,AV}} - \mathrm{LD}^{\mathrm{RP}}] \\
\mathrm{RI}^{\mathrm{test}} = \boldsymbol{R}[\mathrm{C}^{\mathrm{CG/RG,AV}} + \mathrm{C}^{\mathrm{GES}} - \mathrm{C}^{\mathrm{RP}}] \\
\mathrm{RI}^{\mathrm{ref}} = \mathrm{RI}^{\mathrm{test}}
\end{cases}
\tag{42}
$$

Furthermore, the visualization of qualifying the different CC indices is shown in Fig. 4. It should be noted that CC value will be different with respect to different CC indices, however, EPSC are more significant and convincing for the future decarbonized power system [35], and they are which will be specially addressed in the case study.

## III. NUMERICAL ANALYSIS

### A. Set-up

CC evaluation is tested in a standard reliability test system, i.e., IEEE RTS-79 benchmark system, which is shown in Fig. 6. The peak load of the system is 2850 MW, and the total capacity of generations is 3405 MW. Compared with the original system, we further introduce the RG, ES, and VES into the system while maintaining the network topology unchanged. The RG resources account for part of generations and are located at the same buses of the CG, i.e., PV buses. For ES bundled with RG (ES$^{\mathrm{R}}$), ES resources are located with RG, while for distributed ES (ES$^{\mathrm{D}}$), ES resources are located at load buses, i.e., PQ buses. And VES resources stem from load and are located at load buses, i.e., PQ buses. The reliability data of generations and lines are inherited from the original system. The optimization is coded in MATLAB with the YALMIP interface and solved by GUROBI 9.5 solver. The programming environment is Core i7-1165G7 @ 2.80GHz with RAM 16 GB.

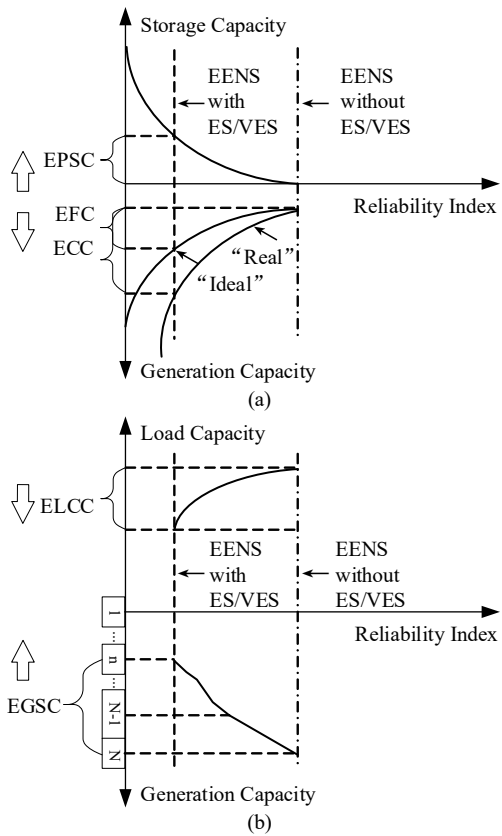

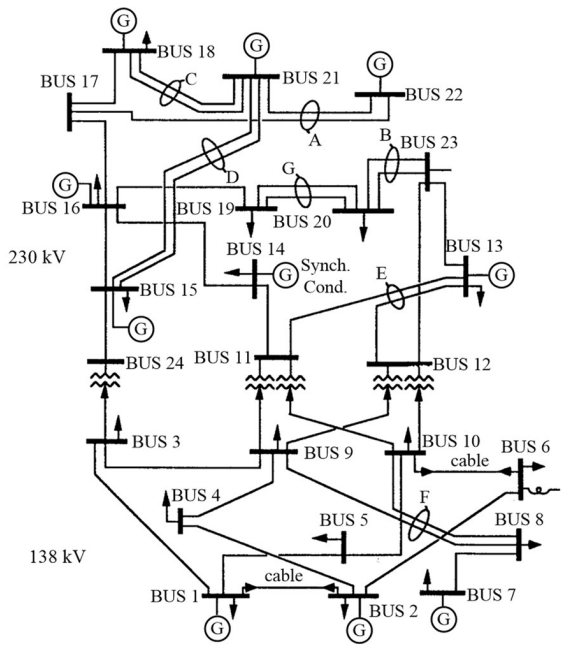

Fig. 5. Visualization of capacity credit (a) EFC, ECC, and EPSC (b) EGCS and ELCC

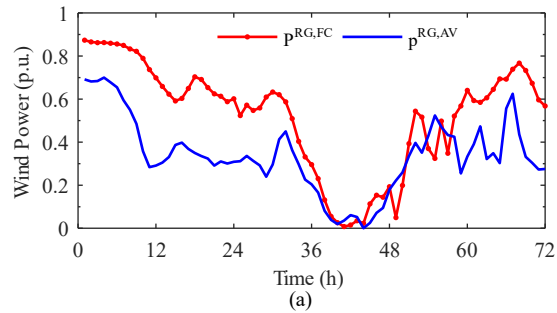

Fig. 6. IEEE RTS-79 benchmark system

## B. Capacity credit evaluation of energy storage

In this section, we compare the performance of the proposed method with two state-of-the-art methods: (i) fixed dispatch and (ii) greedy management. Historical data of RG, load, and day-ahead price are collected from Belgium's transmission system operator, i.e., Elia [36]., and the normalized value of RG and load profile are used as inputs for the test system. The three consecutive days' operation profile is actively demonstrated in Fig. 7. The huge fluctuations of RG and price have been observed in Elia, and local load normally holds two peaks (10 am-2 pm and 6 pm-8 pm). Thus, it is significant for the utilization of ES to reduce peak stress and power fluctuation.

Fig. 7. Operation profile of three typical days for (a) day-ahead forecasted and real-time wind power, (b) real-time load power, (c) day-ahead price

### (a) System and storage operation

EENS is chosen as the referenced reliability index in SMCS for all the examined ES configurations and dispatch methods, because compared with frequency and comprehensive indices, EENS represents the quantity of reliability losses and is more stable. Fig. 8 shows the convergence performance of SMCS and EENS reaches its peak at the beginning, following an immediate reduction at the end of the $1^{st}$ year and a stable convergence at the end of the $5^{th}$ year. We employ the result of the $5^{th}$ year for the following research. The key parameters of Li-ion battery and RG are listed in Table II. In this case, the rated power of ES is noted as the normalized value based on RG capacity, and $ES^R$ is applied for illustration. Fig. 9 compares the great difference in the system and ES operation in two consecutive days, applying the proposed method and the above two alternatives. Table III summarizes the comparison of operations. It is observed that the fixed dispatch method only addresses the system capacity inadequacy due to two load peaks, and the ES follows the "charge-discharge-charge" operation profiles for each day's operation. The fixed dispatch fails to adjust the ES operation to real-time failures (20h-22h and 32h-34h) nor various DIUs and DDUs, though part of the operation follows the day-ahead self-energy management, which results in the maximum energy not served (ENS): 3640 MWh and the minimal flexibility utilization. While greedy management maintains the whole capacity at the normal system state and the ES is discharged in response to the system capacity inadequacy due to any failures. And it outperforms the other two methods with respect to loss of load (ENS: 3127 MWh) and ES follows the "discharge-charge" operation profiles for each day's operation. Although it manages to deal with real-time failures, it overlooks the self-consumption and uncertainties within ES operations. Compared with the above dispatch methods, the proposed method is more complex and realistic, which involves both day-ahead self-energy management (light-blue and light-orange barplot) and real-time adjustment (difference between solid and dotted lines) to real-time failures. The operations of ES consider all the system capacity inadequacy events and it will be recharged after the emergency state, but the adequacy performance (ENS: 3288 MWh) is a little worse than greedy management since the available capacity of ES is subject to the baseline consumption and various DIUs & DDUs.

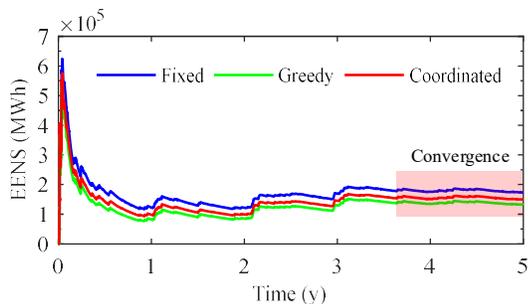

Fig. 8. Convergence performance of SMCS with three dispatch methods

TABLE II KEY PARAMETERS OF ES AND RG

| Parameters | Value | Parameters | Value |
|---|---|---|---|
| $SoC_{i,0}^{ES}$ | 0.4 | $P_{c/d,i,max}^{ES}$ | 30% of RG Capacity |
| $S_i^{ES,rated}$ | 4 h*Rated Power | $\eta_{c/d,i}^{ES}$ | 0.9 |
| $FOR_i^{ES}$ | 5% | $C^{RG,rated}$ | 30% Generation Capacity |
| $\kappa$ | $N(0.5,0.1)$ | $L^{SoH_{instal}-SoH_{end}}$ | 4000 |
| $SoH_{initial}$ | 1.0 | $SoH_{end}$ | 0.8 |
| $SoC_{i,min/max}^{ES}$ | 0.0/1.0 | $\varepsilon_i^{ES}$ | 0.05 |

TABLE III OPERATION PERFORMANCE COMPARISON

| Dispatch Method | Adjustment to RT Failures | DA Self-energy Management | DIUs DDUs | ENS (MWh) |
|---|---|---|---|---|
| Fixed Dispatch | × | ~√ | × | 3640 |
| Greedy Management | √ | × | × | 3127 |
| Coordinated Dispatch | √ | √ | √ | 3288 |

(b) Adequacy performance

Moreover, we compare the adequacy performance among three representative methods and changed with the rated power (10%-50%) and capacity (2h-6h) of ES. Fig. 10 actively demonstrates the adequacy gaps over different operation conditions. It is obvious that the adequacy results follow the order: greedy management > coordinated dispatch > fixed dispatch. And the EENS is reduced with the increasing power and energy capacity, while compared with the energy factor, the increase in rated power accounts for a substantial part of reliability improvement. Moreover, EENS is more sensitive to greedy management and coordinated dispatch, i.e., 10000-20000 MWh/10% rated power and 5000-14000MWh/10% rated power, respectively. While the changes in EENS are relatively small with fixed dispatch, i.e., 1000-4000 MWh/10% rated power. However, the existing methods only generate the theoretical results, but they will not be realized in real-time operations, and it will be proved in the following.

(c) Benefits and limitations of coordinated dispatch

In this part, the benefits and limitations of the application of coordinated dispatch are proved from (i) practical reliability performance, (ii) economic performance, and (iii) safety.

(i) At first glance, the performance of coordinated dispatch is much worse than the greedy management, but it is not true in real-time operations. The practical reliability performance of fixed dispatch and greedy management will be worse than the theoretical ones since they overlook (1) failures in ES, (2) energy losses from self-discharge, and (3) capacity degradation. The consequence of the above aspect is calculated with the 4-h ES operation and shown in Fig. 11 (a), that a considerable reliability reduction is witnessed due to the energy losses from self-discharge, which accounts for 80-90% of the total reliability reduction. And the additional practical reliability losses are increased with the rated power. Additionally, the practical and theoretical reliability performance is summarized in Table VI. It is observed that with the increase in rated power, the practical reliability losses are witnessed as 0.6%, 2.7%, and 8.9% increase for fixed dispatch, while the greedy management bears a larger increase of 1.7%, 5.6%, and 10.7%. Moreover, the practical and theoretical reliability performances are shown in Fig. 10 with dashed and solid lines, respectively. It is observed that the practical performance deviates from the theoretical one, and this gap is apparently increased with increased power and energy capacity. This demonstrates that the results of the existing methods are not realistic, and the increased penetration of ES will worsen this gap and bring more risks due to this overestimation. Thus, the practical reliability performance verifies the benefit of the treatment to DIUs and DDUs of the proposed method.

TABLE VI PRACTICAL AND THEORETICAL ADEQUACY PERFORMANCE OF ES

| Dispatch Method | Performance | Rated Power (4h duration) | | |
|---|---|---|---|---|
| | | 10% | 30% | 50% |
| Fixed Dispatch | Practical | 179375 | 174608 | 175030 |
| | Theoretical | 178269 | 166567 | 160741 |
| Greedy Management | Practical | 162838 | 139521 | 124380 |
| | Theoretical | 160116 | 132136 | 112384 |
| Coordinated Dispatch | Practical | 167682 | 149399 | 136527 |

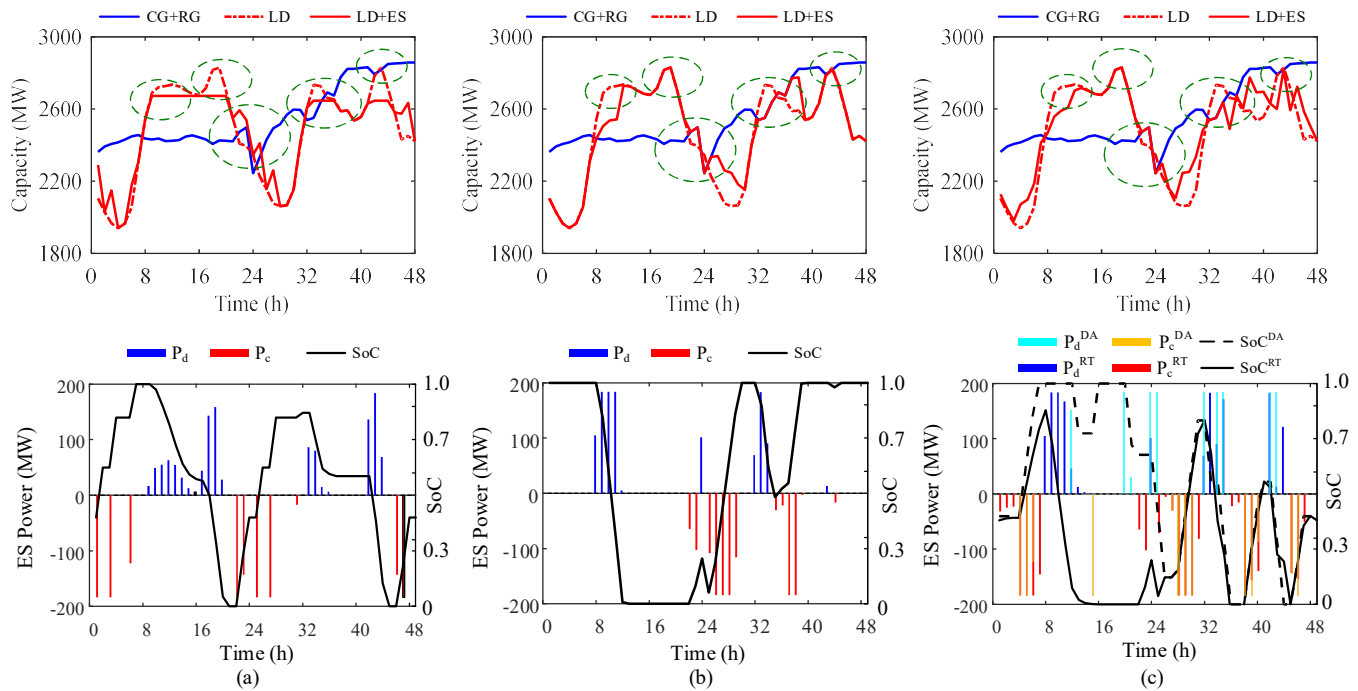

Fig. 9. System and ES operations compared among three methods: (a) fixed dispatch, (b) greedy management, (c) coordinated dispatch

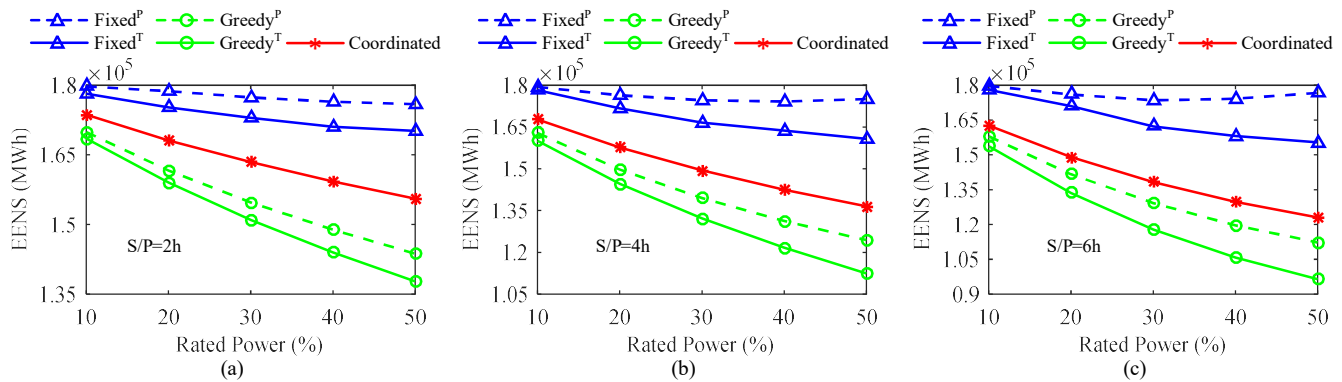

Fig. 10. Adequacy results compared among three methods and rated power & capacity of ES: (a) 2h, (b) 4h, (c) 6h

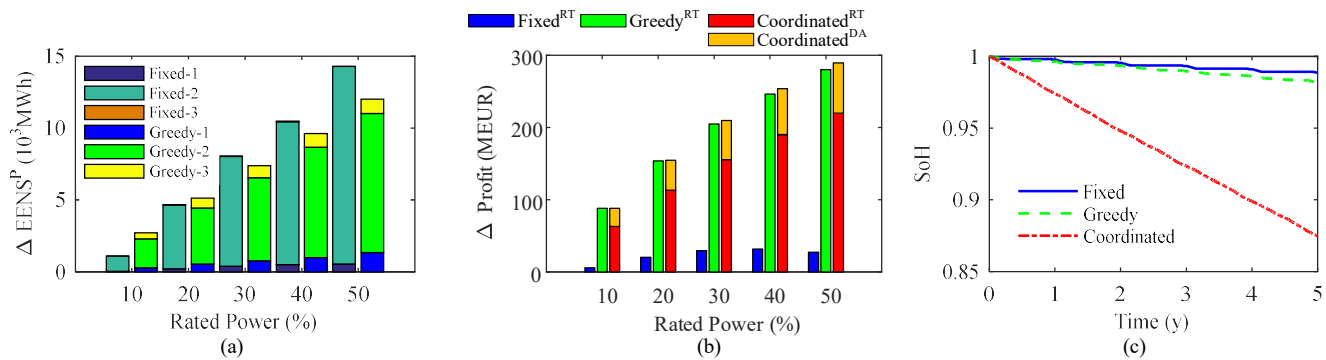

Fig. 11. Benefit and limitation of the proposed methods with respect to (a) practical adequacy performance, (b) economic performance, and (c) safety

(ii) We further compare the day-ahead and real-time economic performance over these three methods, and the results are shown in Fig. 11 (b). The day-ahead profit comes from the energy arbitrage, while adequacy support brings real-time profit. The price of reliability improvement is set to be about 10000 € /MWh in Elia. It is found that both the greedy management and proposed dispatch share a large amount of profit from adequacy support since they provide treatment to real-time failures. Furthermore, only the proposed dispatch has the additional profit from the energy arbitrage, resulting in a better economic performance than greedy management. And economic performance verifies the benefit of treatment to real-time failures and self-energy management. It is worth mentioning that under a more complex market operation, the greedy ES market player will optimally allocate their capacity share for participation in competitive markets [37], which will create more profit from various grid services and this part of the research will be investigated in the future.

(iii) Safety is another significant consideration within ES operations, in this manuscript, the state of health (SoH) is defined as the available energy capacity, while the thermal and electro-chemical risks are not considered. The SoH of ES is calculated with the modeling of capacity degradation, i.e., Eq. (4-5). And the changes in SoH over 5 years of simulation with the alternative dispatch methods are shown in Fig. 11 (c). The stable and slow reduction in SoH is observed with two existing methods, while the dramatic decline in SoH is witnessed with the proposed methods, it is since that the proposed method will frequently switch strategies between real-time and day-ahead one. Although it is a limitation of the proposed method, the operation results will happen in practice and ES players are reluctant to keep their flexible resources idle to respond to the grid even under the risk of battery aging and replacement.

### (d) Capacity credit

In this part, the comparison between different CC indices is analyzed with different ES configurations. Fig. 12 provides a comparison of the existing CC indices with coordinated dispatch. The normalized value based on the rated power of ES is used for illustration. And it is observed that although value difference exists for different CC indices, they share the similarity of declining capacity value with increased penetration of ES as pointed out in [38], i.e., the ability of equivalent capacity replacement will be decreased (from 20%-50% to 15%-40%) with higher penetration of ES power. This actively demonstrates that the investment for ES construction is not quite economic and efficient for the system with higher penetration of ES. And a significant improvement (from 15%-25% to 35%-50%) has been witnessed in response to the increase in energy capacity, which indicates that ES with higher energy capacity is more welcomed for system adequacy. Moreover, speaking of the difference in the CC indices, the CC indices follow the order of EGCS>ECC>EFC>ELCC. And EGCS is more applicable for power system planning since it represents the replacement of generations, while EFC lacks practicality because the perfectly ideal generators do not exist.

Furthermore, EGCS is chosen as the representative for the comparison between different dispatch methods. It is shown in Fig. 13 that ES under greedy management shares the biggest CC value (25%-85%) while the credible capacity is underestimated under fixed dispatch, resulting in 5%-10% normalized EGCS. By contrast, the coordinated dispatch manages to achieve the trade-off between baseline consumption and adequacy support, producing a more realistic value that is nearly half of that of greedy management. Moreover, we compare the ECGS with the proposed method and the existing works, summarized in Table V. Although systematic differences exist in these studies, the CC level and regularity can be widely extended. Ref. [10] employs fixed dispatch in the IEEE-RTS 79 system and considers the changes in power and energy ratings, while the failures in ES and self-discharge are overlooked, thus resulting in overestimated results (10%-45%) compared with its practical performance (5%-22%). Greedy management is used in a real-world transmission system [38] that ES will simply discharge as part of load curtailment if loss of load events happen, while it overlooks the uncertainties and configurations of ES, thus the EGCS (30%-80%) keeps the consistency with changeable ES capacity and is higher than the practical value (26%-83%). While the results in the distribution system [39] shares the similarity of the proposed method since it considers the uncertainties within ES operation, however, the underlying dispatch strategy is extremely simplified.

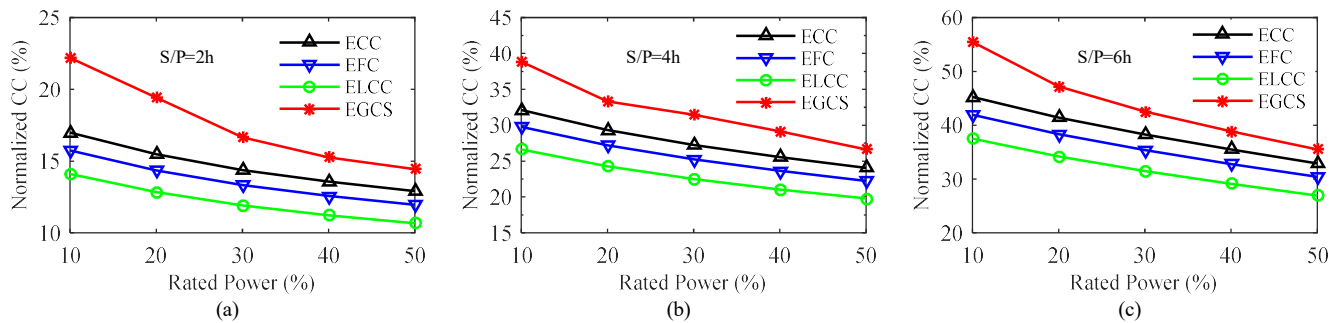

Fig. 12. Normalized Capacity credit value with different indices and power & capacity ratings of ES: (a) 2h, (b) 4h, (c) 6h

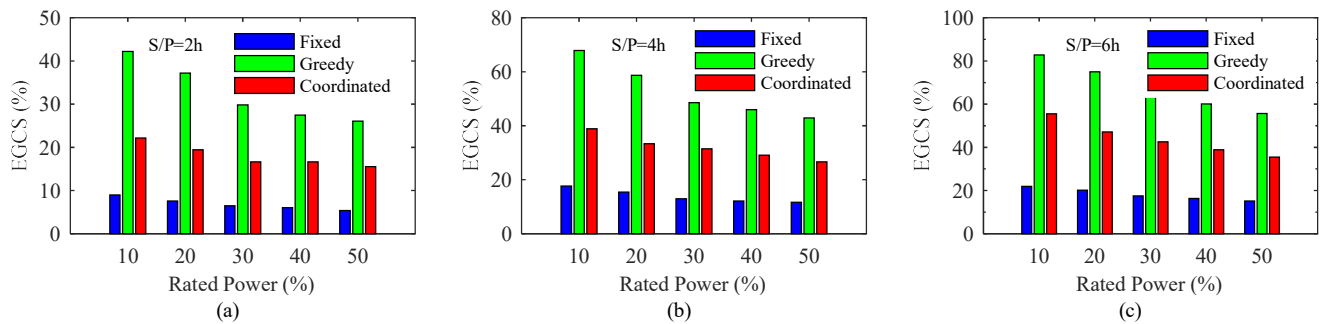

Fig. 13. EGCS compared with different dispatch methods and power & capacity ratings of ES: (a) 2h, (b) 4h, (c) 6h

TABLE V NORMALIZED CAPACITY VALUE OF ES COMPARED WITH STATE-OF-THE-ART METHODS

| Dispatch Method | Energy Capacity | | |
|---|---|---|---|
| | 2h | 4h | 6h |
| Fixed Dispatch | 5%-9% | 12%-18% | 15%-22% |
| Greedy Management | 26%-42% | 42%-68% | 56%-83% |
| Coordinated Dispatch | 16%-22% | 27%-39% | 36%-55% |
| [10] | 10%-20% | 25%-40% | 35%-45% |
| [38] | | 30%-80% | |
| [39] | 15%-20% | 30%-40% | 40%-55% |

### C. Capacity credit evaluation of virtual energy storage

While different from ES, we address the CC evaluation of VES in this subsection. The impact of the dispatch method is hard to examine since the $VES^T$ mainly contributes to the peak load. Thus, the greedy management is unrealistic (infeasible to charge/discharge except on peak load days) and the coordinated dispatch has a slight advantage over the fixed dispatch (considering the adjustment of failures during peak load days). However, the uncertainties within the dispatch affect much of the performance. Therefore, we compared the result of dispatch (U1) without uncertainties modeling, (U2) considering DIUs from self-consumption, and (U3) considering both DIUs and DDUs from self-consumption and human behavior. The data inputs of load and $VES^T$ are derived from real households in the Muller project in Austin, TX, USA [40], where $VES^T$ accounts for a substantial part of load and parameters of $VES^T$ model are generated by parameter identification and model aggregation [32]. The visualization of uncertainties and variations of $VES^T$ operations are shown in Fig. 14 concerning load power, on-state probability, $VES^T$ baseline consumption, and SoC bounds under DIUs, where boxplots represent the variations across different weeks and red lines represent the mean value of the probabilistic parameters. It is observed that $VES^T$ tends to operate at a lower consumption level at late night and in the morning, but it is relatively higher in the evening (after working and back home). Also, the SoC level of $VES^T$ tends to be around 0.5 at night to maintain a comfortable temperature for sleep, but an obvious increase happens in the afternoon and evening to maintain a cool temperature for work.

#### (a) System and storage operation

To keep the consistency with the evaluation of ES, EENS is also chosen as the referenced reliability index in SMCS for all the examined VES configurations and dispatch methods. And we also employ the simulation result of the $5^{th}$ year for the following research. The key parameters of $VES^T$ and RG have listed in Table VI. In this case, the rated power of $VES^T$ is noted as the maximum normalized value based on Load capacity. Compared with ES, $VES^T$ generally embraces a higher self-discharge rate and lower energy capacity. And the model parameters are time-varying and subject to self-consumption.

Fig. 15 compares the considerable difference in the system and VES operation on a typical day, applying the proposed methods with different uncertainties. It is observed that U1 outperforms the others in residual capacity contribution and VES flexibility utilization, it is since that the constant and full available SoC/ energy capacity is assumed within the model, resulting in an over-estimated result (ENS: 2923 MWh). However, in consideration of 1) DIUs from self-consumption behavior and 2) DDUs from incentive and discomfort, a sharp decline is witnessed in the available SoC/energy capacity (40% and 50% decline on peak load time, respectively.) as long as the VES power actions (50% and 60% decline on peak load time, respectively). This leads to worse results of the loss of load, i.e., 3509 and 3534 MWh, respectively. Although the consideration of uncertainties worsens the performance, the operations under DIUs and DDUs reveal the practical actions of VES resources, which is more credible and realistic.

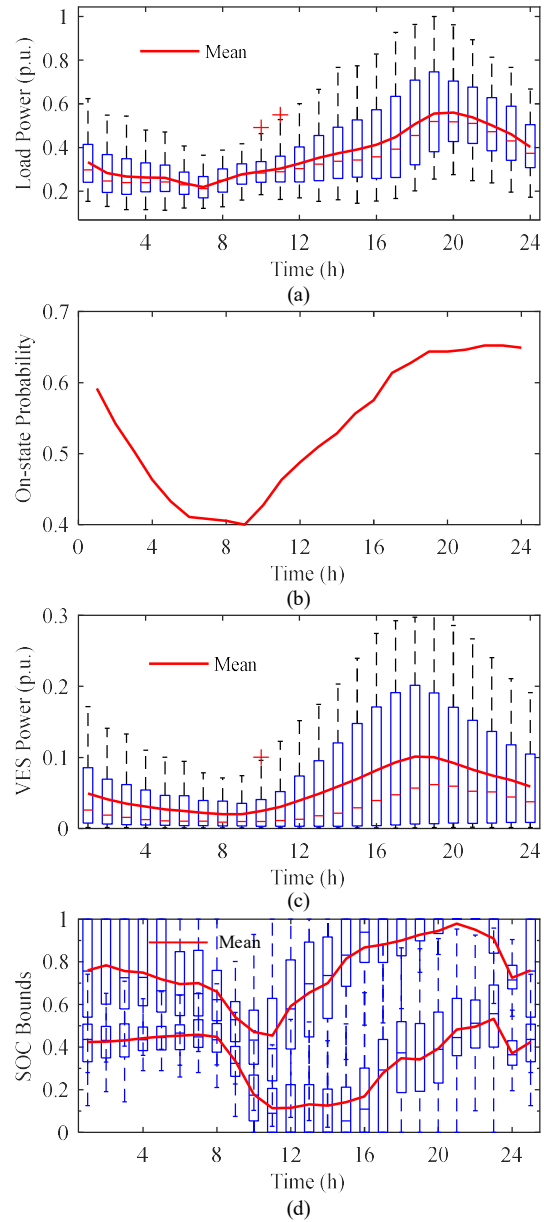

Fig. 14. Visualization of uncertainties in VES operations with respect to: (a) load power, (b) on-state probability, (c) baseline consumption, (d) SoC bounds

TABLE VI KEY PARAMETERS OF VES AND RG

| Parameters | Value | Parameters | Value |
|---|---|---|---|
| $SoC_{i,0}^{VES}$ | 0.6 | $P_{c/d,i,max}^{VES}$ | 30% of Load Capacity |
| $S_i^{VES,rated}$ | 1 h*Rated Power | $\eta_{c/d,i}^{VES}$ | 1 |
| $SoC_{i,min/max}^{VES}$ | 0.0/1.0 | $\varepsilon_i^{VES}$ | 0.5 |
| $C^{RG,rated}$ | 30% Generation Capacity | $a_g^- / a_{\underline{g}}$ | 1/1 |
| $\rho$ | 0.5 | $b_{\overline{h}}^- / b_{\underline{h}}$ | 2/6 |
| $C_{c/d}^{VES}$ | 50/250 | $\overline{C}^{VES}$ | 300 |

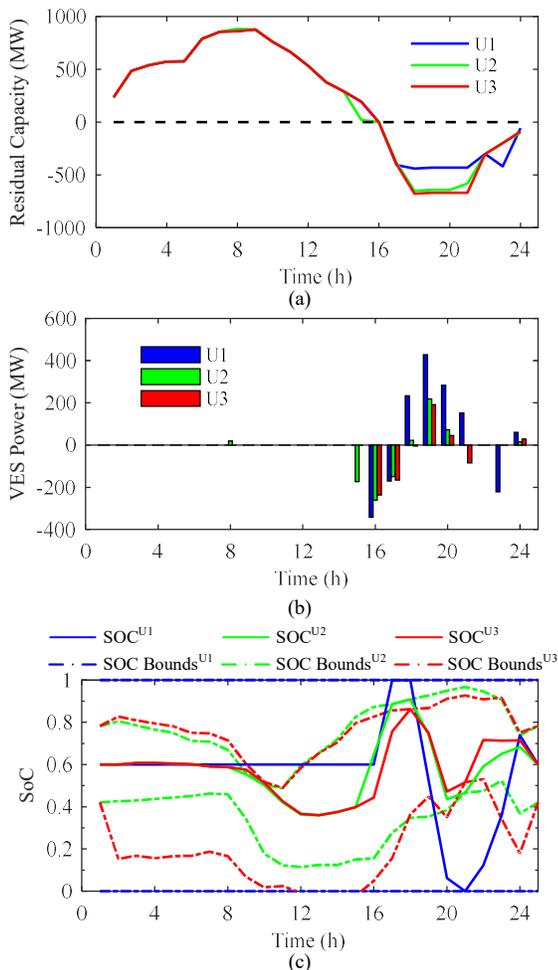

Fig. 15. Comparison of system and VES operations: (a) residual capacity, (b) VES Power, and (c) SoC and SoC bounds

(b) Adequacy performance

Moreover, we compare the adequacy performance among different uncertainties and changed with the rated power (10%-50%) and capacity (1h-2h) of VES. Fig. 16 actively demonstrates the adequacy difference over different operation conditions. It is observed that the adequacy results follow the order: U1 > U2 > U3, which is in line with the order of available SoC bounds of VES. And similar to ES performance, the EENS is reduced with the increasing power and energy capacity, and compared with the energy factor, the increase in rated power accounts for a substantial part of reliability improvement. Moreover, EENS reduction is more sensitive with U1 at 2000-3000 MWh/10% rated power compared with U2 and U3 at 1000-2000 MWh/10% rated power.

However, the theoretical results will not be realized in real-time operations, since it overlooks DIUs or DDUs within operations, resulting in the response unavailability (during DR) and load rebound (after DR). Thus, in this paper, we follow the DIUs and DDUs structure discussed above, to calculate the response deviation with different uncertainty assumptions. The performance of 50% rated power and 2h duration VES is shown in Fig. 17. It is observed that U1 and U2 have huge response losses during peak load time (blue bar), while few response losses are witnessed in U3 (red bar).

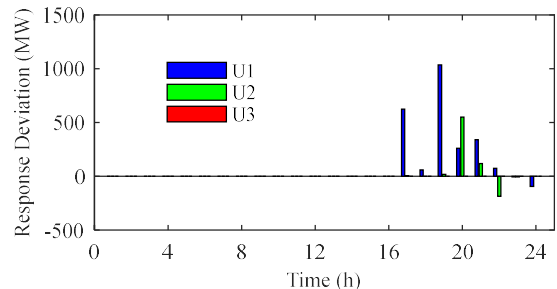

Fig. 17. Response deviation of the proposed method with different uncertainties (50% rated power 2h duration VES)

In addition, the practical adequacy performance is marked with dashed lines in Fig. 16. It is observed that the practical EENS greatly deviates from the theoretical ones in U1, while a slight increase and decline are witnessed in U2 and U3. Moreover, the consequence of DIUs and DDUs (deviation) has increased with the increased power and energy capacity. For small capacity VES, operations under DIUs and DDUs have a slight inferiority with respect to adequacy performance. However, for large capacity VES, the operations under DIUs and DDUs are credible though conservative, we summarize the operations of 2h VES concerning theoretical and practical performance in Table VII. It should be mentioned that the consequence of uncertainties has a larger impact on the adequacy performance of VES than that of ES because the daily capacity decline is revealed for VES compared with the annual capacity degradation for ES. Moreover, compared with the theoretical result, the practical one of U1 and U2 increased by 20%-100% and 5%-25%, respectively. While barely difference in EENS is witnessed for U3, it is since that the chance-constrained method under DDUs manages to control the response unavailability of GES within the security level (1-γ). Thus, we can conclude that modeling DIUs and DDUs of VES is highly required for CC evaluation of large capacity VES (e.g., over 30% rated power and 2h duration). And finding solutions to reduce the conservation of U3 will be discussed in the following.

TABLE VII PRACTICAL AND THEORETICAL ADEQUACY PERFORMANCE OF VES

| Dispatch Method | Performance | Rated Power (2h duration) | | |
|---|---|---|---|---|
| | | 10% | 30% | 50% |
| U1 | Practical | 25022 | 23341 | 22903 |
| | Theoretical | 20730 | 13389 | 11444 |
| U2 | Practical | 26029 | 22305 | 19715 |
| | Theoretical | 24555 | 18864 | 15695 |
| U3 | Practical | 25249 | 19983 | 16787 |
| | Theoretical | 25249 | 19983 | 16787 |

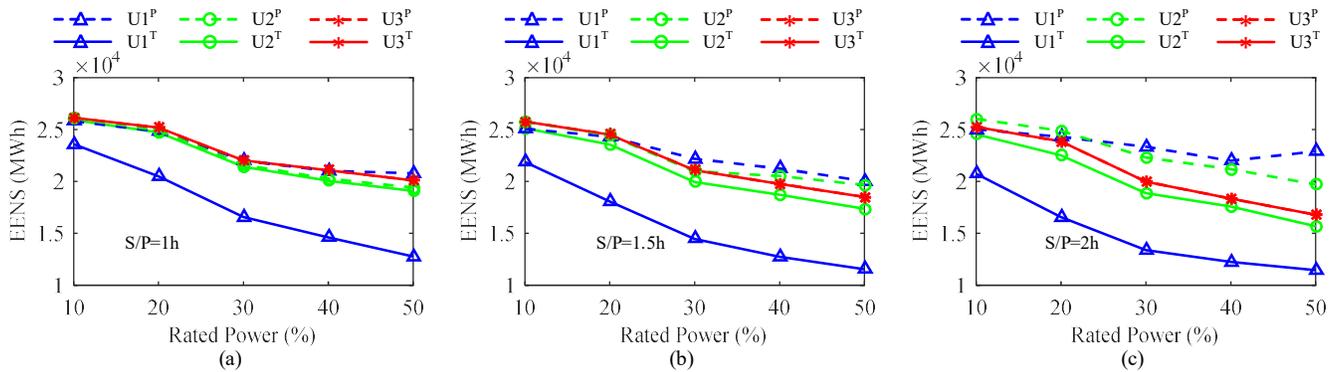

Fig. 16. Adequacy results compared among different uncertainties and rated power & capacity of VES: (a) 1h, (b) 1.5h, (c) 2h

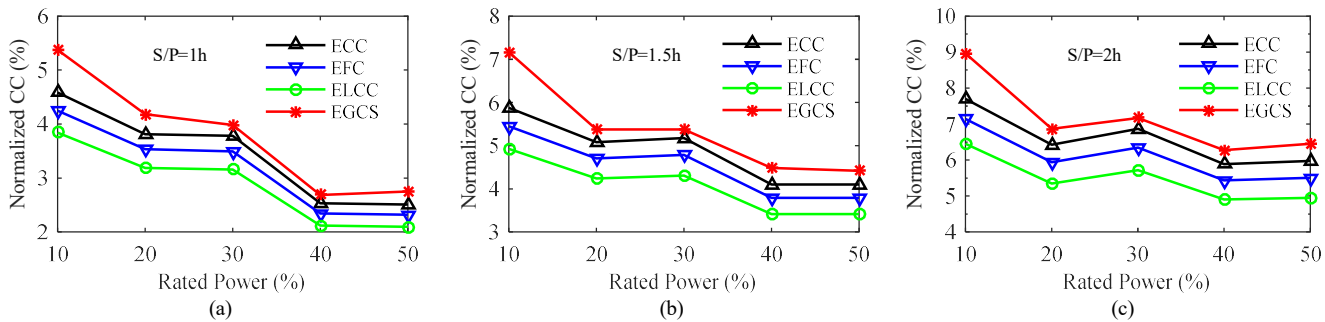

Fig. 18. Normalized Capacity credit value with different indices and power & capacity ratings of VES: (a) 1h, (b) 1.5h, (c) 2h

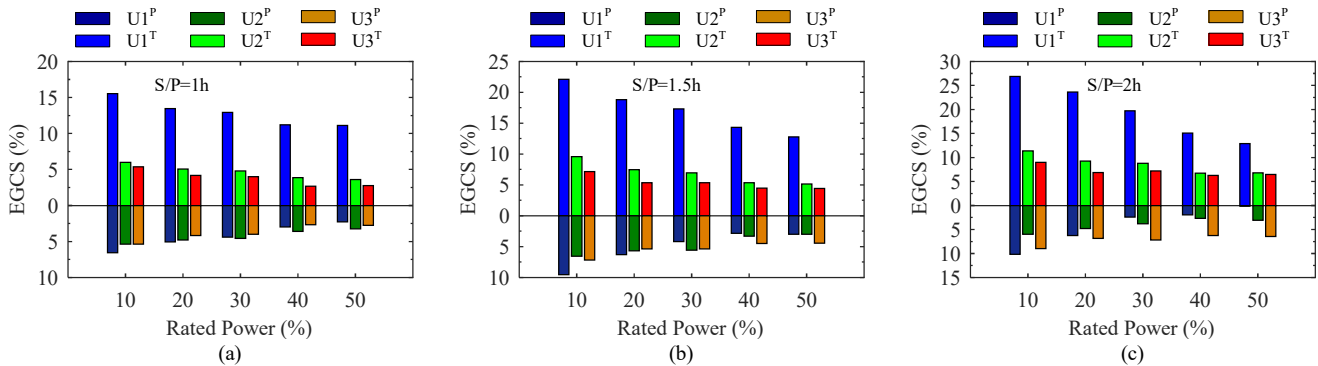

Fig. 20. EGCS compared with different uncertainties and power & capacity ratings of ES: (a) 1h, (b) 1.5h, (c) 2h

### (d) Capacity credit

In this part, the comparison between different CC indices is analyzed with different VES configurations and uncertainties. Fig. 18 provides a comparison of the existing CC indices with the proposed method (U3). The normalized value based on the rated power of VES is used for illustration. And it is observed that although value difference exists for different CC indices, they share the similarity of declining capacity value with increased penetration, i.e., the ability of equivalent capacity replacement will be decreased (from 5.4%-9.0% to 2%-6.0%) with higher penetration of VES power, while this decline is not apparent compared with ES. And for high-capacity VES, CC value witnesses a slight increase with increased rated power, as pointed out in [25]. Moreover, speaking of the difference in the CC indices, the CC indices follow the order of EGCS>ECC>EFC>ELCC. And the CC value of VES is relatively lower than that of ES, it is since that VES mainly accounts for the loss of load events during peak load time.

In addition, the capacity of VES in equivalence with ES is investigated and evaluated by the proposed EPSC. The comparison results of 2h-duration VES with different uncertainties and rated power are shown in Fig. 19. This demonstrates that the ability of physical storage replacement is increased with the rated power of VES. Without uncertainty consideration, the VES is equivalent to 60%-95% of ES, while the practical DIUs and DDUs extremely reduce this ability, resulting in 25%-60% EPSC in practice.

Furthermore, EGCS is chosen as the representative for the comparison between different dispatch methods. It is shown in Fig. 20 that the over-estimated theoretical CC value (11%-27%) is witnessed without uncertainty modeling, though resulting in the relatively low practical CC value (1%-10%). By contrast, considering DIUs and DDUs manages to reduce the gap between the theoretical and practical CC value. Moreover, we compare the ECGS with the proposed method and existing works and they are summarized in Table VIII. Although systematic differences exist in these studies, the CC level and regularity can be widely extended. It is observed that the consequence of DIUs and DDUs will change the theoretical regularity, resulting in a sharp decline in practical EGCS for high-capacity VES. And CC with U3 outperforms the others, especially for larger power and energy capacity. Ref. [10] employs fixed dispatch of VES and the theoretical result (20%-30%) is based on ideal and constant VES, which is in line with the theoretical result of U1. While considering DIUs and DDUs, the practical CC shares the similarity of U3, i.e., 9%-11%. Another relevant work [25] investigates the impact of both the DR unavailability and load rebound impact on CC value. Although a considerable decline from 26%-46% to 16%-39% is observed with DDUs impact, the CC value is much higher than the others. It is since that the VES resources are assumed to be available every day, while the mean load factor for VES is 10%-15% used in this paper, thus VES only accounts for less than 5% of the load peak on average. And with 5% rated power, VES only performs at 8% and 17% for practical and theoretical CC values [25], which are similar to the result of U1 and U3 and verify the validity of the proposed method.

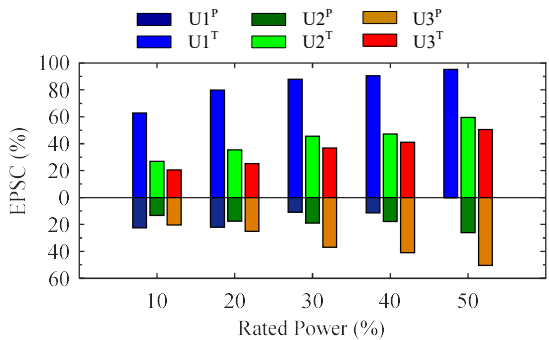

Fig. 19. EPSC compared with different uncertainties and rated power of VES

TABLE VIII Normalized Capacity Value of VES compared with State-of-the-Art Methods

| Dispatch Method | Performance | Energy Capacity | | |
|---|---|---|---|---|
| | | 1h | 1.5h | 2h |
| U1 | Practical | 2.7%-6.0% | 3.2%-9.6% | 1.0%-10.7% |
| | Theoretical | 11.1%-15.6% | 12.8%-22.1% | 12.9%-22.9% |
| U2 | Practical | 3.2%-5.4% | 3.0%-6.6% | 3.1%-6.0% |
| | Theoretical | 3.6%-6.0% | 5.1%-9.6% | 6.8%-11.4% |
| U3 | Practical | 2.7%-5.4% | 4.4%-7.2% | 6.5%-9.0% |
| | Theoretical | 2.7%-5.4% | 4.4%-7.2% | 6.5%-9.0% |
| [10] | Practical | | 9%-11% | |
| | Theoretical | | 20%-30% | |
| [25] | Practical | | 16%-39% | |
| | Theoretical | | 25%-46% | |

## IV. Application of Physical and Virtual Energy Storage in Decarbonized Power Systems

Decarbonization of the electricity sector is one of the major measures in slowing down the pace of climate change. And the expansion of variable renewable energy resources has become the main solution for transitioning towards a low-carbon energy supply and realizing the Paris Agreement and decarbonization goals of each country [41]. However, the inherent stochasticity within the RES operations causes enormous challenges for power system operations, especially under high-penetration scenarios. Recent works have pointed out that energy storage plays a significant role in the application of high penetration of RES with respect to cost, RES curtailment, $CO_2$ emission, etc [42]-[43]. While few pieces of work address the capacity value of storage in decarbonized power systems. Thus, in this section, the numerical analysis demonstrates the enhancement of both the adequacy of the decarbonized power system with the support of ES and VES under different affecting factors. Moreover, suggestions for capacity market and generation replacement are further provided. Since the coordinated dispatch method outperforms the state-of-the-art methods, the following statement is only based on the proposed method.

### A. Key factors affecting the adequacy performance with ES

In this subsection, we investigate the adequacy performance of ES with different affecting factors: (a) power and capacity ratings of ES, (b) charge/discharge efficiency of ES, (c) MTTR of energy storage, and (d) location of ES.

(a) Power and capacity ratings of ES

The configuration of energy storage is the most important factor that affects the adequacy enhancement of ES. Different power and capacity ratings of ES have been employed in the decarbonized power system with different penetration of RES and the adequacy performance is shown in Fig. 21 and Fig. 22. The capacity value is summarized in Table IX. It is observed that the adequacy performance will be improved by both the increased power and capacity ratings, and the capacity rating becomes a more prominent contributor. Moreover, the reduction of EENS is increasing with the penetration of RES under the Low-RES scenario (0%-30%), and the upward increase becomes gradually stable even with a little decline under the Medium-RES scenario (40%-60%), then following by a slow-growing performance under High-RES scenario (70%-100%). This actively demonstrates that it is more efficient to employ energy storage under the Low-RES scenario, while it is suggested to employ other flexibilities for power systems with higher penetration RES. Additionally, the EGCS witnesses a slight decline under the Low-RES scenario and a sharp decline under the Medium-RES scenario, following a slow increase under the High-RES scenario. This essentially confirms the above conclusion and verifies that the ability of generation replacement is decreased for power systems with low and medium penetration of RES. Although the above finding is independent of power and energy ratings, long-time duration energy storage outperforms the other configuration of ES and achieves up to 67% capacity value, which provides a possible solution for low-carbon systems in the future.

TABLE IX Normalized Capacity Value compared with RES Penetration and Energy Capacity of ES

| Penetration | Energy Capacity | | |
|---|---|---|---|
| | 2h | 4h | 6h |
| Low (0%-30%) | 16%-33% | 25%-67% | 35%-67% |
| Medium (40%-60%) | 7%-22% | 10%-45% | 15%-56% |
| High (40%-60%) | 7%-23% | 10%-43% | 15%-56% |

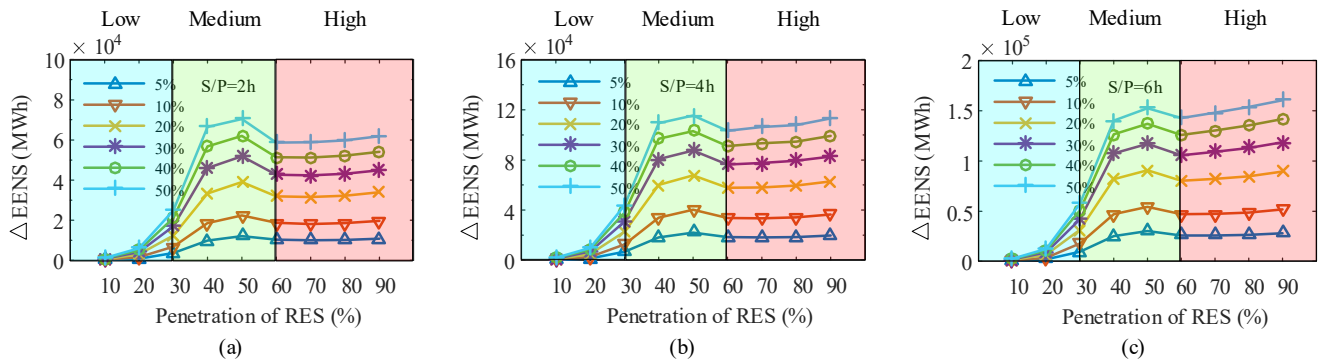

Fig. 21. EENS changes of decarbonized power system with different penetration of RES and power & capacity ratings of ES: (a) 2h, (b) 4h, (c) 6h

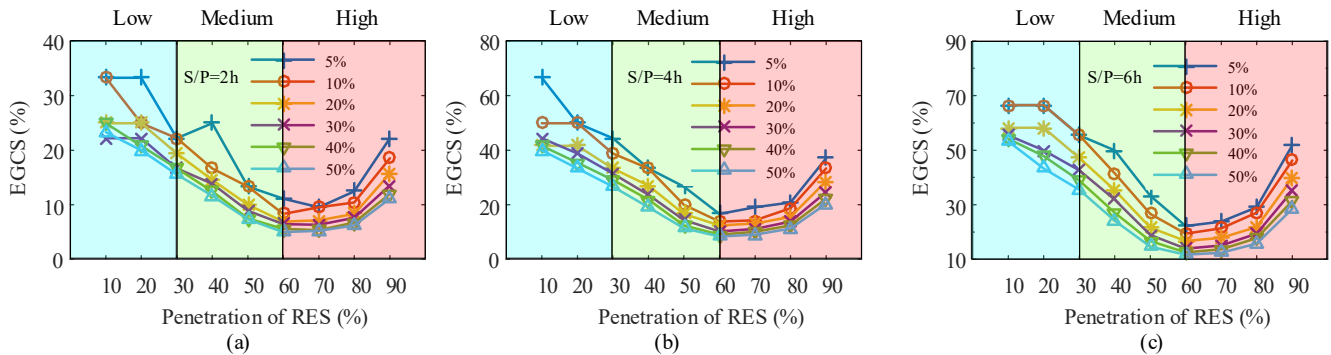

Fig. 22. EGCS of decarbonized power system with different penetration of RES and power & capacity ratings of ES: (a) 2h, (b) 4h, (c) 6h

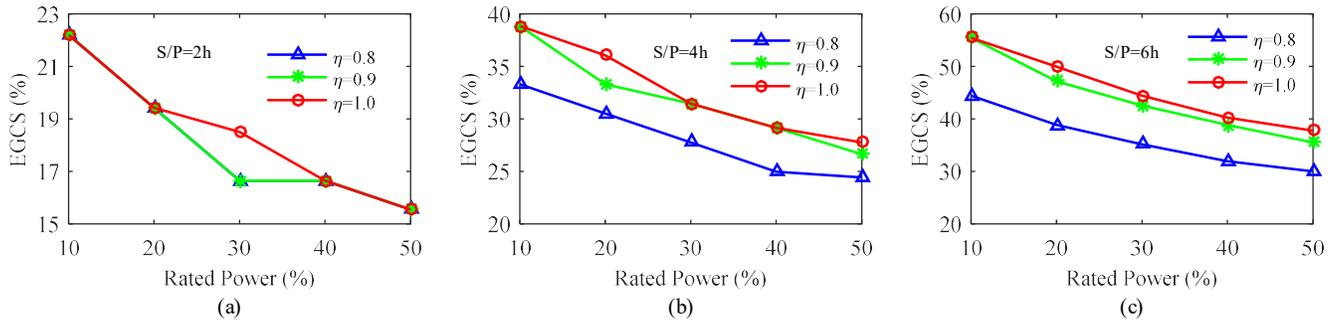

Fig. 23. EGCS of decarbonized power system with different efficiency and power & capacity ratings of ES: (a) 2h, (b) 4h, (c) 6h

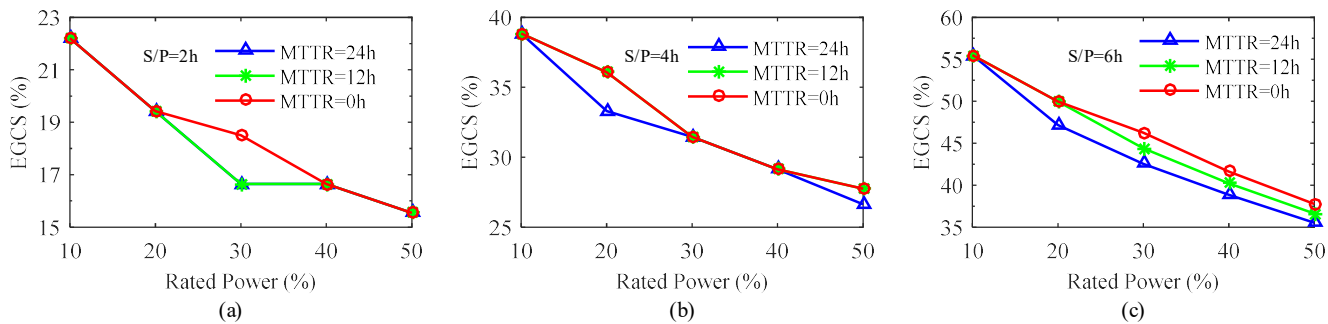

Fig. 24. EGCS of decarbonized power system with different MTTR and power & capacity ratings of ES: (a) 2h, (b) 4h, (c) 6h

(b) Charge/discharge efficiency ES

Charge/discharge efficiency is another key factor affecting the operations of ES and its capacity value. In this part, we compare the ES with different types, including low-efficiency ES (compressed air energy storage), high-efficiency ES (Li-ion battery,) and ideal ES. The comparison of the normalized capacity value of ES with efficiency and power & capacity ratings is listed in Table X and vividly shown in Fig. 23. The capacity value is computed with the 30% penetration of RES. It can be seen that with short-time duration ES, the efficiency difference has little impact on its capacity value (only 2% gaps for 30% rated power). However, gaps in the adequacy contribution become larger for long-time duration energy storage (10% gaps for 6-h ES). This demonstrates that it is of great significance to improve the efficiency of long-time duration ES for the decarbonized power system.

TABLE X Normalized Capacity Value compared with Efficiency and Energy Capacity of ES

| Efficiency | Energy Capacity | | |
|---|---|---|---|
| | 2h | 4h | 6h |
| $\eta = 0.8$ | 15%-22% | 25%-33% | 33%-45% |
| $\eta = 0.9$ | 15%-22% | 27%-39% | 40%-56% |
| $\eta = 1.0$ | 15%-22% | 28%-39% | 43%-56% |

(c) MTTR of ES

We further investigate the impact of the failure property of ES and we employ ES with different MTTR, i.e., 24h, 12h, and 0h (ideal ES). The comparison of capacity value is shown in Fig. 24 and summarized in Table XI. It is observed that the failure property of ES mainly impacts the long-time duration ES, and the difference gaps can reach up to 5%. Thus, reducing the time to repair is beneficial to operations with long-time duration ES.

TABLE XI Normalized Capacity Value compared with MTTR and Energy Capacity of ES

| MTTR | Energy Capacity | | |
|---|---|---|---|
| | 2h | 4h | 6h |
| 24h | 15%-22% | 27%-39% | 36%-56% |
| 12h | 15%-22% | 28%-39% | 37%-56% |
| 0h | 15%-22% | 28%-39% | 38%-56% |

(d) Location of ES

To find the better allocation of ES, compared with the ES bundled with RG, we employ the distributed ES that ES resources are more distributed at load buses. The adequacy improvement is shown in Fig. 25. It is obvious that the EENS is a little reduced with small power and capacity ratings (200MWh for 30%-40% 4h duration $ES^D$), however, the adequacy is less improved for large capacity configuration. It is since that $ES^D$ directly reacts to the local capacity inadequacy while avoiding the energy loss of the network, however, the ES with large capacity overcome the limitations of the network.

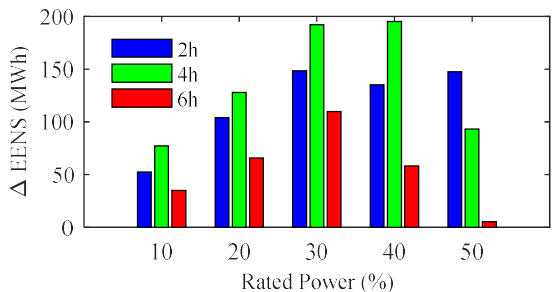

Fig. 25. Adequacy improvement with distributed energy storage with different power and capacity ratings

## B. Key factors affecting the adequacy performance with VES

In this subsection, we investigate the adequacy performance of VES with different affecting factors: (a) power and capacity ratings of VES, (b) self-discharge rate of VES, (c) load factor and correlation of VES, (d) probabilistic level of chance-constrained method, and (e) DDU level.

(a) Power and capacity ratings of VES

The configuration of storage is the most important factor that affects the adequacy enhancement of VES. While different from ES, the capacity of VES is limited by the thermodynamics of buildings ($VES^T$) or the latency time of customers ($VES^E$). Different power and capacity ratings of VES have been employed in the decarbonized power system with different penetration of RES and the adequacy performance is shown in Fig. 26. It is observed that the adequacy performance will be improved by both the increased power and capacity ratings, and the capacity rating becomes a more prominent contributor. Compared with ES, VES with low capacity has relatively worse even "0" performance. Moreover, the reduction of EENS is increasing with the penetration of RES under the Low-RES scenario (0%-30%), and the increase becomes gradually stable even with a sharp decline under the Medium-RES scenario (40%-60%), then following by a stable or slight decrease under High-RES scenario (70%-100%). This actively demonstrates that it is more efficient to employ VES under the Low-RES scenario, and low-capacity VES fails to support power systems with higher penetration RES.

Additionally, the capacity value is summarized in Table XII. And Fig. 27 illustrates that the EGCS witnesses a slight increase under the Low-RES scenario and a sharp decline under the Medium-RES scenario, and EGCS equals "zero" with over 50% penetration of RES. This essentially confirms the above conclusion and verifies that VES's ability of generation replacement is only applicable with low and medium penetration of RES. Moreover, long-time duration VES still outperforms the other configuration and achieves up to 16.1% capacity value, which is 80%-90% equivalent to ES, thus finding long-duration VES resources is essential for the decarbonized power system.

TABLE XII Normalized Capacity Value compared with RES Penetration and Energy Capacity of VES

| Penetration | Energy Capacity | | |
|---|---|---|---|
| | 1h | 1.5h | 2h |
| Low (0%-30%) | 5.4%-6.0% | 4.4%-13.1% | 6.5%-16.1% |
| Medium (40%-60%) | 0%-1.8% | 1.1%-2.4% | 1.8%-4.2% |
| High (40%-60%) | 0%-0% | 0%-1.1% | 1.2%-3.0% |

(b) Self-discharge rate of VES

Compared with ES and $VES^E$, $VES^T$ embraces a relatively larger self-discharge rate. which has a direct relationship with building parameters, as shown in Eq. (43). $R$, $C$, and $K$ are the thermal resistance, thermal capacity, and conversion efficiency of $VES^T$. In this part, while maintaining the other factors unchanged, we investigate the adequacy performance with

$$
\begin{cases}
\varepsilon^{\mathrm{VES}} = 1 - e^{-\Delta t / RC} \\
S^{\mathrm{VES}} = \dfrac{\Delta t (T_{\max}^{\mathrm{in}} - T_{\min}^{\mathrm{in}})}{KR\varepsilon^{\mathrm{VES}}} \\
P^{\mathrm{VES}} = \dfrac{T^{\mathrm{out}} - T^{\mathrm{in}}}{KR}
\end{cases}
\tag{43}
$$

different self-discharge rates (i.e., 0.3,0.5, and 0.7), which is vividly illustrated in Fig. 28. The result of EGCS is summarized in Table XII. It is observed that VES with a larger self-discharge rate outperforms the others and the EGCS increase from 2.7%-7.8% ( $\varepsilon = 0.3$ ) to 4.2%-11.4% ( $\varepsilon = 0.7$ ). And this gap is increased with increased power and energy capacity. However, the regularity is the opposite for ES. It is since that the adequacy performance with ES will be declined with increased self-discharge losses of energy, while the self-discharge losses are manifested in temperature for VES. And according to Eq. (43), a larger self-discharge rate represents a larger temperature tolerance range or smaller thermal resistance. Thus, it is suggested to find VES resources with a larger temperature tolerance range or smaller thermal resistance.

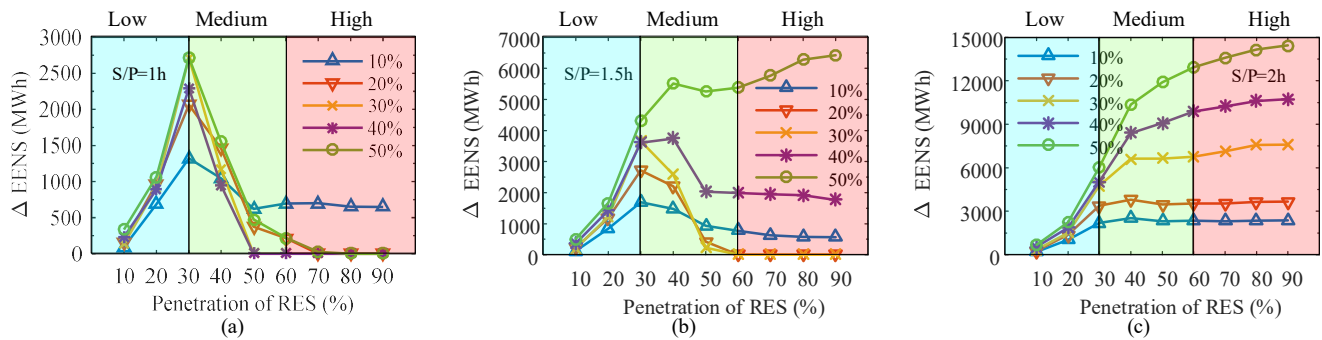

Fig. 26. EENS changes of decarbonized power system with different penetration of RES and power & capacity ratings of VES: (a) 1h, (b) 1.5h, (c) 2h

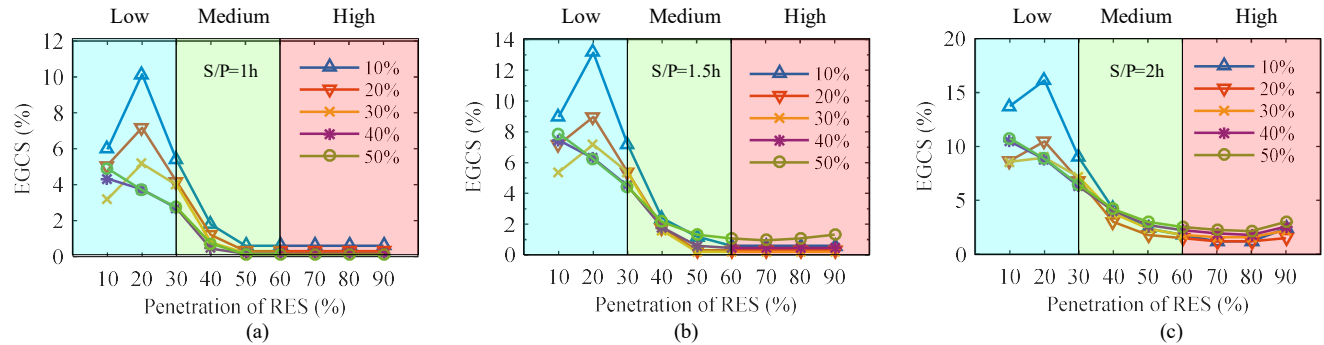

Fig. 27. EGCS of decarbonized power system with different penetration of RES and power & capacity ratings of VES: (a) 1h, (b) 1.5h, (c) 2h

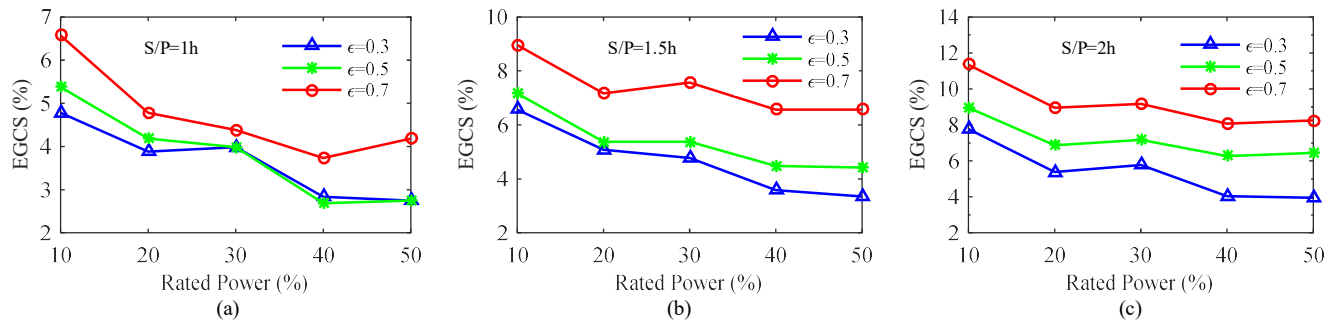

Fig. 28. EGCS of decarbonized power system with different self-discharge rate and power & capacity ratings of ES: (a) 1h, (b) 1.5h, (c) 2h

TABLE XII Normalized Capacity Value compared with Self-discharge Rate and Energy Capacity of VES

| Self-discharge Rate | Energy Capacity | | |
|---|---|---|---|
| | 1h | 1.5h | 2h |
| $\varepsilon = 0.3$ | 2.7%-4.7% | 3.3%-6.6% | 3.9%-7.8% |
| $\varepsilon = 0.5$ | 2.7%-5.4% | 4.4%-7.2% | 6.5%-9.0% |
| $\varepsilon = 0.7$ | 4.2%-6.6% | 6.6%-9.0% | 8.2%-11.4% |

(c) Load factor and correlation factor of VES

In the above analysis, the CC value of VES is quantified with the load data derived from TCL. However, in reality, as individuals may have different energy consumption patterns, the adequacy contribution of VES could vary tremendously with different load characteristics. In this part, we focus on the load factor and correlation factor and investigate their impact on CC value by four different types of VES. 1) The first type of VES is derived from household electric equipment except for heating or cooling, e.g., lighting, oven, etc. This kind of load accounts for baseline consumption with a high load factor and correlation. 2) The second type of VES is TCL from air-conditioning which is assumed in the above case and only works during summer days. This kind of load has a relatively low load factor. 3) Compared with VES$^2$, VES$^3$ is assumed to work throughout the whole year (e.g., located with high temperatures all year round). And both the load factor and load correlation are higher than VES$^2$. 4) The fourth type of VES is from EV load, and its load profile is shifted 5 hours later than VES$^3$. This kind of load maintains the same load factor of VES$^3$ but with a relatively low correlation factor. The EGCS is calculated with 2h VES with the above four types and summarized in Table XII. Noted that VES with higher load factor and higher correlation factor outperforms the others in the ability of adequacy contribution. This demonstrates that it is significant to check the load characteristic of VES. And for VES resource targeting, it is effective to target VES which accounts for a substantial part of load shape and load level.

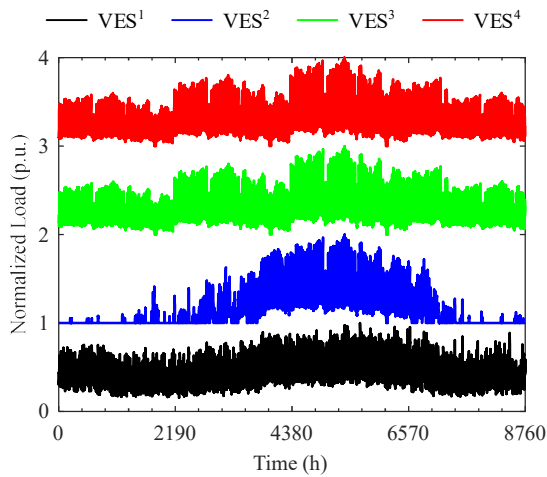

Fig. 29. Load profile of VES resources with different load patterns

TABLE XIII NORMALIZED CAPACITY VALUE COMPARED WITH LOAD FACTOR AND CORRELATION FACTOR OF VES

| VES Type | Load Factor | Correlation Factor | EGCS |
|---|---|---|---|
| VES$^1$ | 0.45 | 0.85 | 15.2%-22.4% |
| VES$^2$ | 0.17 | 0.74 | 6.5%-9.0% |
| VES$^3$ | 0.33 | 0.94 | 35.8%-40.6% |
| VES$^4$ | 0.33 | 0.15 | 10.4%-15.2% |

(d) Probabilistic level of chance-constrained method

The set of probability levels is a tradeoff between the DDUs' consequence and the flexibility utilization of VES resources. Also, the probability level limits both the response reliability of VES and the reliability of the system. Fig. 30 illustrates the adequacy improvement and EGCS with different probability levels. It is obvious that though the reliability improvement is increased with the probabilistic level, while the changes in improvement are stable and even with a slight decline for 2h-duration VES after 35%. It is since that flexibility utilization increases with probabilistic level and DDUs' consequences do not affect much on the overall system adequacy at first. However, when the probability level continues to increase over

35%, the DDUs' consequence becomes serious and will reduce the system's reliability. Thus, the decision-maker is suggested to achieve the best tradeoff at around 35% probability level.

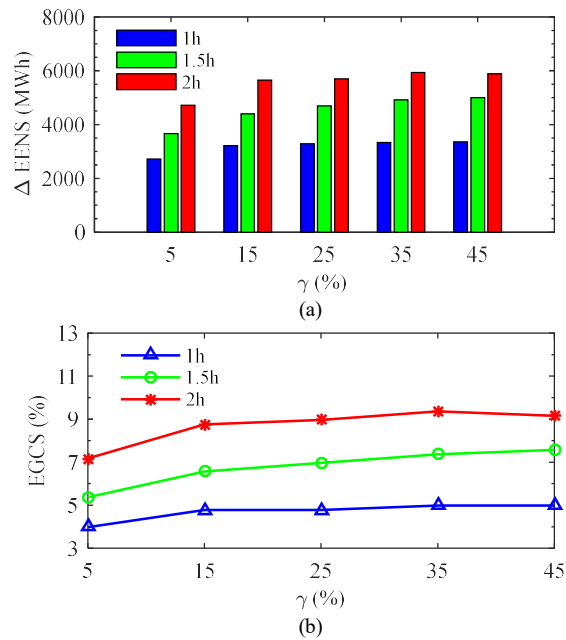

Fig. 30. The impact of the probabilistic level of chance-constrained method (a) adequacy improvement, (b) EGCS

(e) DDUs level

In addition to the probability level, the DDUs level is another important factor to be considered. Herein, the probability level is guaranteed to be constant at 5%, and we change the incentive effect and discomfort effect by changing the $a_g$ and $b_h$ in the DDUs structure. It should be noted that the incentive effect and discomfort effect are two opposite DDU effects that will affect the available SoC bounds/available capacity of VES. The setup in Section III. B is used as the baseline DDUs level, and the incentive and discomfort aversion are changed by 0.5-1.5 times of the baseline level. The 2h-duration VES is employed as an illustration and the results are summarized in Table XIV. Both the increase in the incentive level and the decrease in the discomfort level will benefit the performance of VES, while the results are more sensitive to the discomfort level. Specially, EGCS is increased by 29.2%-35.6% with a 50% decrease in discomfort level. By contrast, EGCS has only increased by 10.8%-17.8% with a 50% increase in incentive level. This indicates that it is more profitable to target VES resources with lower discomfort aversion) rather than invest more in incentives.

TABLE XIV NORMALIZED CAPACITY VALUE COMPARED WITH DDUs LEVEL OF VES

| Incentive Level | Discomfort Level | EGCS |
|---|---|---|
| 1 | 1 | 6.5%-9.0% |
| 0.5 | 1 | 4.1%-7.2% |
| 1.5 | 1 | 7.2%-10.6% |
| 1 | 0.5 | 8.4%- 12.2% |

## V. Status Quo and Suggestions for Capacity Market

In addition to the scientific research described in the above discussion, in this section, we will introduce the status quo and suggestions for real-world capacity market operations.

### A. Status quo and regulations

The capacity market is designed as part of the wholesale market in many countries (e.g., UK, USA, Ireland, etc.) [17] [44]-[45], which ensures long-term grid reliability by procuring the appropriate amount of power supply resources needed to meet predicted energy demand. Capacity remuneration mechanisms ensure an adequate level of capacity market participation, where the market players are paid by CC in advance, and the final settlement is determined by the practical performance that the players will get rewards for over-performance or be penalized for under-performance. Therefore, CC is not only an important indicator for the system security threshold but also has a significant impact on the capacity market operation and pricing mechanism to avoid overestimation or underestimation of the actual capacity value that should be remunerated.

However, different from generators, the existing regulations for CC qualification of ES and VES are insufficient and immature. Table XV provides the regulations for qualifying CC in some of the regional transmission organizations (RTOs and independent system operators (ISOs). It is observed that the CC of storage is qualified with simple and fixed regulations in most ISOs. For instance, ISO-New England (ISO-NE) and CAISO required the minimum discharge duration of storage facilities and used a derating capacity factor to qualify CC. Storage with 2h, 4h, and 6h discharge duration are qualified to be 50%, 100%, and 100% CC in CAISO. Furthermore, a piecewise CC factor method is proposed by New York Independent System Operator (NYISO) to qualify storage resources according to the rated power and maximum discharge duration. While Pennsylvania-New Jersey-Maryland Interconnection (PJM) applies a more advanced approach for ES and VES, i.e., using ELCC to evaluate the capacity substitution of the generator, however, this method has limitations in huge computing pressure with a large number of storage participants. Additionally, it is actively demonstrated that CC of VES is not defined in some ISOs, and others just follow the same approach of ES, which all overlook the inherent DIUs and DDUs within the resources.

TABLE XV  Comparisons of Capacity Credit Regulation in Different Capacity Markets

| ISOs/RTOs | BES | VES |
|---|---|---|
| PJM | ELCC | ELCC |
| ISO-NE | Capacity Factor-2h | Not Defined |
| MISO | Not Defined | Not Defined |
| NYISO | Piecewise Capacity Factor | Not Defined |
| CAISO | Capacity Factor-4h | Capacity Factor-4h |

### B. Capacity value compared with academic works

The capacity value of ES employed in capacity markets are listed in Table XVI, as a comparison with the academic works. It is found that although certain difference exists in different market and transmission system, the capacity value follows the regularity of the previous works, that the CC of ES is increased with energy capacity. And 6h-duration ES generally has a full capacity value. However, the capacity values assumed in the current capacity markets are overestimated and the results are in line with the performance of greedy management (2h: 26%-42%, 4h: 42%-68%, 6h: 56%-83%). Thus, the CC values listed in Table XVI are only applicable for credible capacity bidding that ES is able to deliver adequacy support without any self-consumption or other market behaviors.

TABLE XVI  Comparisons of Derating Factors of Storage in Different Capacity Markets

| ISOs/RTOs | Energy Capacity | | |
|---|---|---|---|
| | 2h | 4h | 6h |
| CAISO | 50% | 50% | 100% |
| NYISO | 45% | 90% | 100% |
| Elia | 56% | 79% | 100% |
| EIRGRID | 33.6%-39.6% | 60.5%-53.6% | 77.0%-68.1% |
| National Grid ESO | 39.7% - 49.2% | 64.9% - 71.4% | 94.6% |

In addition, we focus on the capacity value of DR in CAISO. The real data of seven representative days in August and September of the year 2020 are chosen as an illustration in Fig. 31. It is observed that about one-third of the capacity requirement provided by DR is not available or directly accessible in real-time dispatch, though about one-sixth of the capacity is provided in excess of the theoretical capped capacity value (yellow bar). Additionally, the availability performance drops significantly on weekends and holidays, also at the start and the end of DR duration, which may be affected most by energy consumption behavior. The capacity credit (dash-dotted line) appears to be 30% over-counted compared to the practical contribution (blue and yellow bar), which mainly lies in the simplified modeling of uncertainties from customer behaviors.

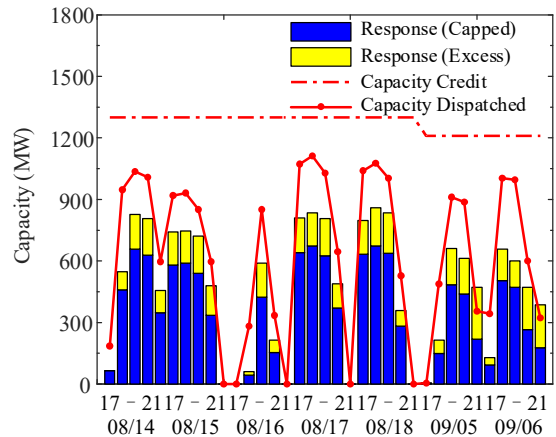

Fig. 31. Capacity performance of DR in CAISO

*C. Suggestions*

Noted that the current regulations of the capacity markets are over-simplified and over-optimistic, which will bring huge gaps in theoretical and practical capacity performance. Thus, to achieve a more reasonable and realistic capacity value, suggestions for capacity market operations are listed as follows.

- It is suggested to incorporate both the human and market behavior of ES/VES into CC evaluation, thus the learning of baseline consumption and real-time performance is of vital significance.

- Employ/target long-time duration ES is a feasible and significant solution for the decarbonized power system. While VES is subject to its short-time duration, which is only applicable for systems with low penetration of RES.

- Although the CC of ES/VES will be altered with the dispatch method, uncertainties, system configuration, etc., some regularity can be widely extended. For instance, CC will be declined with increased rated power capacity and increased with increased rated energy capacity. And under the same rated power and capacity, the CC of VES is generally lower than that of ES.

- Power & energy capacity and charge/discharge efficiency are key factors affecting the CC of ES. While CC of VES is more sensitive with load factor, correlation factor, and DDUs level.

- Ensuring timely CC calculation is highly required for capacity market operations. Thus, acceleration methods, marginal CC, and storage aggregators are suggested to improve computational efficiency. And derating factors of CC can be learned with both the historical and simulation-based performance of ES/VES.

## VI. Conclusion

In this paper, we present a novel methodological framework for qualifying the CC of ES/VES in the future decarbonized power system. Compared with the previous works, the primary novelty of the proposed framework is to incorporate modeling of market and human behavior of ES/VES into CC evaluation. A two-stage coordinated dispatch is proposed to achieve the trade-off between day-ahead market behavior and efficient adjustment to real-time failures. Moreover, different from "ideal" and "constant" storage properties, we further modeled human behavior with storage operations with two types of DIUs and one type of DDUs. Furthermore, novel reliability and CC indices are introduced to evaluate the practical and theoretical adequacy contribution of ES and VES. Exhaustive case studies verify the following conclusions:

1) The proposed method manages to generate a credible and realistic CC value of ES/VES and simulated with dynamic and available capacity of storage, while the previous works either overestimate or underestimate the CC value with simplified modeling of storage.

2) The CC value of ES is highly affected by the dispatch method, while the uncertainty modeling of storage contributes

a substantial part of the variety in CC value of VES. And overlooking DIUs and DDUs within the dispatch will cause huge consequences (10%-70% overestimated CC).

3) CC value of ES/VES is not only dependent on the power and energy ratings. Charge/discharge efficiency, load and correlation factor, and DDUs level are also key factors affecting their adequacy performance.

4) Adequacy performance of ES/VES will be changed with different decarbonized levels, and long-time duration & high-reliability storages are a sustainable solution for the future power system.

Future research will focus on some practical challenges that remain unaddressed. In particular, the presented framework will be extended to incorporate a cost-benefit analysis of ES/VES. And apart from their adequacy performance, the economic performance compared with generation investment will be investigated. And as mentioned above, employing acceleration methods and learning the adequacy performance from historical data will be further discussed.

**Credit authorship contribution statement**

Ning Qi: Conceptualization, Methodology, Software, and Writing. Peng Li: Funding Support. Lin Cheng: Supervision. Ziyi Zhang: Funding Support. Wenrui Huang: Format Review and Revision. Weiwei Yang: Funding Support.

**Declaration of Competing Interest**

The authors declare that they have no known competing financial interests or personal relationships that could have appeared to influence the work reported in this paper.

**Data availability**

Please refer to Mendeley Dataset where we upload all the raw data and processed data used for this paper [47]. And we are much grateful to the data provider: Pecan Street and ENTSO-e.

**Acknowledgments**

This paper was sponsored by the National Key R&D Program of China (Grant No. 2021YFB2400700), Scientific research project of China Three Gorges Corporation (Grant No. 202203028), National Natural Science Foundation of China (Grant No. 52037006). We would like to give our special appreciation to Prof. Pierre Pinson (Imperial College London) and Prof. Mads. R. Almassalkhi (University of Vermon) who gave significant support to this paper during Ning Qi's visit to Denmark.

**Ning Qi** received a B.S. degree in electrical engineering from Tianjin University, China, in 2018. He is currently pursuing a Ph.D. degree in electrical engineering with the State Key Laboratory of Control and Simulation of Power Systems and Generation Equipment, Department of Electrical Engineering, Tsinghua University, China. His research interests include data-driven and optimization methodologies and their applications to generic energy storage and virtual power plant.

**Peng Li** received a Master's degree in electrical engineering. He is currently a senior engineer at China Three Gorges Renewables. His research interests include the grid-aware dispatch of renewable energy and energy storage.

**Lin Cheng** received a B.S. degree in electrical engineering from Tianjin University, China, in 1996 and received a Ph.D. degree from Tsinghua University, China, in 2001. He is currently a tenured professor at the Department of Electrical Engineering, Tsinghua University. He is also a fellow of *IET* and a senior member of *IEEE*. His research interests include operational reliability evaluation and application of power systems, operation optimization of distribution systems considering flexible load-side resources, and perception and control of uncertainty in wide-area measurement systems.

**Ziyi Zhang** received a Master's degree in electrical engineering. He is currently an engineer at China Three Gorges Renewables. His research interests include the grid access of renewable energy and transmission system planning.

**Wenrui Huang** received a Master's degree in electrical engineering. He is currently an engineer at the State Key Laboratory of Control and Simulation of Power Systems and Generation Equipment, Department of Electrical Engineering, Tsinghua University, China. His research interests include the operation, maintenance, and evaluation technology of electrochemical energy storage.

**Weiwei Yang** received a Master's degree in electrical engineering. He is currently an engineer at China Three Gorges Renewables. His research interests include power system simulation and energy storage technology.

## REFERENCES

[1] Allan, R. N. Reliability evaluation of power systems. Springer Science & Business Media.

[2] IEA, "Global energy review 2021." [Online]. Available: https://www.iea.org/reports/global-energy-review-2021.

[3] Höök M, Tang X. Depletion of fossil fuels and anthropogenic climate change—A review. Energy policy. 2013 Jan 1;52:797-809.

[4] Khan AS, Verzijlbergh RA, Sakinci OC, De Vries LJ. How do demand response and electrical energy storage affect (the need for) a capacity market?. Applied Energy. 2018 Mar 15;214:39-62.

[5] Lata P, Vadhera S. Reliability improvement of radial distribution system by optimal placement and sizing of energy storage system using TLBO. Journal of Energy Storage. 2020 Aug 1;30:101492.

[6] Qi N, Cheng L, Zhuang Y, Zhou Y, Zhang Y, Zhu C. Reliability assessment and improvement of distribution system with virtual energy storage under exogenous and endogenous uncertainty. Journal of Energy Storage. 2022 Dec 1;56:105993.

[7] Haslett J, Diesendorf M. The capacity credit of wind power: A theoretical analysis. Solar Energy. 1981 Jan 1;26(5):391-401.

[8] Garver LL. Effective load carrying capability of generating units. IEEE Transactions on Power apparatus and Systems. 1966 Aug(8):910-9.

[9] Amelin M. Comparison of capacity credit calculation methods for conventional power plants and wind power. IEEE Transactions on Power Systems. 2009 Mar 27;24(2):685-91.

[10] Zhou Y, Mancarella P, Mutale J. Framework for capacity credit assessment of electrical energy storage and demand response. IET Generation, Transmission & Distribution. 2016 Jun;10(9):2267-76.

[11] Safdarian A, Degefa MZ, Lehtonen M, Fotuhi-Firuzabad M. Distribution network reliability improvements in presence of demand response. IET Generation, Transmission & Distribution. 2014 Dec;8(12):2027-35.

[12] Zhou Y, Mancarella P, Mutale J. Modelling and assessment of the contribution of demand response and electrical energy storage to

adequacy of supply. Sustainable Energy, Grids and Networks. 2015 Sep 1;3:12-23.

[13] Evans MP, Tindemans SH, Angeli D. Minimizing Unserved Energy Using Heterogeneous Energy Storage Unit (vol 34, pg 3647, 2019). IEEE Transactions on power systems. 2020 Sep 1;35(5):4144-.

[14] Denholm P, Nunemaker J, Gagnon P, Cole W. The potential for battery energy storage to provide peaking capacity in the United States. Renewable Energy. 2020 May 1;151:1269-77.

[15] ENTSO-E, "European resource adequacy assessment 2021–annex3: Methodology." [Online]. Available: https://extranet.acer.europa.eu/en/Electricity/Pages/European-resource-adequacy-assessment.aspx.

[16] EPRI, "Midwest independent transmission system operator (miso) energy storage study." [Online]. Available: https://www.epri.com/research/products/1024489.

[17] CAISO, "Demand response issues and performance." [Online]. Available: http://www.caiso.com/participate/Pages/Load/Default.aspx

[18] Haessig P, Multon B, Ahmed HB, Lascaud S, Bondon P. Energy storage sizing for wind power: impact of the autocorrelation of day‐ahead forecast errors. Wind Energy. 2015 Jan;18(1):43-57.

[19] Qi N, Pinson P, Cheng L, Wan Y, Zhuang Y. Chance Constrained Economic Dispatch of Generic Energy Storage under Decision-Dependent Uncertainty. arXiv preprint arXiv:2201.06407. 2022 Jan 17.

[20] Bhattarai S, Karki R, Piya P. Reliability and economic assessment of compressed air energy storage in transmission constrained wind integrated power system. Journal of Energy Storage. 2019 Oct 1;25:100830.

[21] Cheng L, Wan Y, Zhou Y, Gao DW. Operational reliability modeling and assessment of battery energy storage based on Lithium-ion battery lifetime degradation. Journal of Modern Power Systems and Clean Energy. 2021 Nov 24.

[22] Moshari A, Ebrahimi A, Fotuhi-Firuzabad M. Short-term impacts of DR programs on reliability of wind integrated power systems considering demand-side uncertainties. IEEE Transactions on Power Systems. 2015 Jul 16;31(3):2481-90.

[23] Nikzad M, Mozafari B. Reliability assessment of incentive-and priced-based demand response programs in restructured power systems. International Journal of Electrical Power & Energy Systems. 2014 Mar 1;56:83-96.

[24] Kwag HG, Kim JO. Reliability modeling of demand response considering uncertainty of customer behavior. Applied Energy. 2014 Jun 1;122:24-33.

[25] Zeng B, Wei X, Zhao D, Singh C, Zhang J. Hybrid probabilistic-possibilistic approach for capacity credit evaluation of demand response considering both exogenous and endogenous uncertainties. Applied energy. 2018 Nov 1;229:186-200.

[26] Allan RN, Billinton R, Lee SH. Bibliography of the application of probability methods in power system reliability evaluation 1977-1982. IEEE Power Engineering Review. 1984 Feb(2):24-5.

[27] Hashem M, Abdel-Salam M, Nayel M, El-Mohandes MT. A Bi-level optimizer for reliability and security assessment of a radial distribution system supported by wind turbine generators and superconducting magnetic energy storages. Journal of Energy Storage. 2022 Jul 1;51:104356.

[28] Hasche B, Keane A, O'Malley M. Capacity value of wind power, calculation, and data requirements: the Irish power system case. IEEE Transactions on power systems. 2010 Aug 16;26(1):420-30.

[29] Milligan M, Frew B, Ibanez E, Kiviluoma J, Holttinen H, Söder L. Capacity value assessments of wind power. Advances in Energy Systems: The Large‐scale Renewable Energy Integration Challenge. 2019 Mar 18:369-84.

[30] Dong W, Chen X, Yang Q. Data-driven scenario generation of renewable energy production based on controllable generative adversarial networks with interpretability. Applied Energy. 2022 Feb 15;308:118387.

[31] Jiang Y, Kang L, Liu Y. A unified model to optimize configuration of battery energy storage systems with multiple types of batteries. Energy. 2019 Jun 1;176:552-60.

[32] Qi N, Cheng L, Xu H, Wu K, Li X, Wang Y, Liu R. Smart meter data-driven evaluation of operational demand response potential of residential air conditioning loads. Applied Energy. 2020 Dec 1;279:115708.

[33] Schneider KP, Sortomme E, Venkata SS, Miller MT, Ponder L. Evaluating the magnitude and duration of cold load pick-up on residential distribution using multi-state load models. IEEE Transactions on Power Systems. 2015 Nov 11;31(5):3765-74.

[34] Nan S, Zhou M, Li G. Optimal residential community demand response scheduling in smart grid. Applied Energy. 2018 Jan 15;210:1280-9.

[35] Misconel S, Zöphel C, Möst D. Assessing the value of demand response in a decarbonized energy system–A large-scale model application. Applied Energy. 2021 Oct 1;299:117326.

[36] Elia. Open access data of Belgium transmission system operator: [Online]. Available: https://www.elia.be/

[37] Yan X, Gu C, Wyman-Pain H, Li F. Capacity share optimization for multiservice energy storage management under portfolio theory. IEEE Transactions on Industrial Electronics. 2018 Mar 22;66(2):1598-607.

[38] Parks K. Declining capacity credit for energy storage and demand response with increased penetration. IEEE Transactions on Power Systems. 2019 May 9;34(6):4542-6.

[39] Konstantelos I, Strbac G. Capacity value of energy storage in distribution networks. Journal of Energy Storage. 2018 Aug 1;18:389-401.

[40] McCracken B, Crosby M, Holcomb C, Russo S, Smithson C. Data-driven insights from the nations deepest ever research on customer energy use. Pecan Res. Inst., Austin, TX, USA. 2013:1949-3053.

[41] United Nation, "Paris Agreement." [Online]. Available: https://unfccc.int/process-and-meetings/the-paris-agreement/the-paris-agreement

[42] Sepulveda NA, Jenkins JD, Edington A, Mallapragada DS, Lester RK. The design space for long-duration energy storage in decarbonized power systems. Nature Energy. 2021 May;6(5):506-16.

[43] Jafari M, Korpås M, Botterud A. Power system decarbonization: Impacts of energy storage duration and interannual renewables variability. Renewable Energy. 2020 Aug 1;156:1171-85.

[44] U.K., "Capacity market." [Online]. Available: https://www.gov.uk/government/collections/electricity-market-reform-capacity-market

[45] Byers C, Levin T, Botterud A. Capacity market design and renewable energy: Performance incentives, qualifying capacity, and demand curves. The Electricity Journal. 2018 Jan 1;31(1):65-74.

[46] Xiao JW, Yang YB, Cui S, Liu XK. A new energy storage sharing framework with regard to both storage capacity and power capacity. Applied Energy. 2022 Feb 1;307:118171.

[47] Supporting data brief [Online]. Available: https://data.mendeley.com/datasets/h5rccz3nw6/draft?a=872f9c09-9ac5-4073-a962-74e4706519b5

[48] Liu, Z., Hou, K., Jia, H., Zhao, J., Wang, D., Mu, Y., & Zhu, L. A Lagrange multiplier based state enumeration reliability assessment for power systems with multiple types of loads and renewable generations. IEEE Transactions on Power Systems, 36(4), 3260-3270.